\documentclass[a4paper,fleqn]{cas-sc}
\usepackage[numbers]{natbib}
\usepackage{graphics}
\usepackage{adjustbox}
\usepackage{textcomp}
\usepackage{lineno}
\usepackage{setspace}
\begin{document}
%\lipsum[1]
\doublespacing
%\setstretch{3.}
\let\WriteBookmarks\relax
\def\floatpagepagefraction{1}
\def\textpagefraction{.001}
\shorttitle{Ageing tests of MSMGRPC in high irradiation dose}
\shortauthors{D. Bartos et~al.}

\title [mode = title]{Ageing studies of Multi-Strip Multi-Gap Resistive Plate Counters based on low resistivity glass electrodes in high irradiation dose}
                     
%\tnotemark[1,2]

%\tnotetext[1]{This document is the results of the research
%   project funded by the National Science Foundation.}

%\tnotetext[2]{The second title footnote which is a longer text matter
%   to fill through the whole text width and overflow into
%   another line in the footnotes area of the first page.}

%\cormark[1]
%\ead{cvr_1@tug.org.in}
%\ead[url]{www.cvr.cc, cvr@sayahna.org}

%\credit{Conceptualization of this study, Methodology, Software}

\address[1]{Hadron Physics Department, National Institute for Physics and Nuclear Engineering - IFIN-HH, P.O. Box MG-6, Bucharest-Magurele, Romania}

\author[1]{D. Bartos}
\author[3]{C. Burducea}
\author[3]{I. Burducea}
\author[1]{G. Caragheorgheopol}
\author[3]{F. Constantin}
\author[3]{L. Craciun}
\author[1]{D. Dorobantu}
\author[5]{M. Ghena}
\author[3]{D. Iancu}
\author[5]{A. Marcu}
\author[1]{G. Mateescu}
\author[3]{P. Mereuta}
\author[2]{V. Moise}
\author[4]{C. Negrila}
\author[2]{D. Negut}
\author[1]{M. Petris}
\author[1]{M. Petrovici}[orcid=0000-0002-7783-9029]
%%\cormark[1]
\author[1]{L. Radulescu}
\author[1]{V. Aprodu}[]
\author[1]{L. Prodan}
\author[1]{A. Radu}
\author[1]{G. Stoian}

\address[2]{Multipurpose IrradiationCentre, National Institute for Physics and Nuclear Engineering - IFIN-HH, P.O. Box MG-6, Bucharest-Magurele, Romania}

\address[3]{Applied Nuclear Physics Department, National Institute for Physics and Nuclear Engineering - IFIN-HH, P.O. Box MG-6, Bucharest-Magurele, Romania}

\address[4]{National Institute of Materials Physics, INCDFM, P.O. Box MG-7, Bucharest-Magurele, Romania}

\address[5]{Institute for Laser, Plasma and Radiation Physics - INFLPR, P.O. Box MG-36, Bucharest-Magurele, Romania}

\cortext[cor1]{Corresponding author: M. Petrovici}

\begin{abstract}
Detailed tests and analysis of ageing effects of high irradiation dose on 
Multi-Strip Multi-Gap Resistive Plate Counters (MSMGRPC) based on low resistivity 
glass electrodes, foreseen to be used for the most forward polar angles covered by the
Time-of Flight (ToF) sub-detector of the Compressed Baryonic Matter (CBM) experiment at 
Facility for Antiprotons and Ion 
Research (FAIR) - Darmstadt are reported. The tests were performed at a multi-purpose irradiation 
facility of IFIN-HH based on $^{60}$Co source.
MSMGRPC efficiency, cluster size, surface and 
volume resistivity of the glass electrodes after 
irradiation are measured and compared with their values before irradiation.
The results of a comprehensive analysis of the composition and properties of the
deposited layers on the glass electrodes, based on different methods, i.e. Scanning Electron 
Microscope (SEM), X-ray Photoelectron Spectroscopy 
(XPS), foil Elastic Recoil Detection Analysis (ERDA), Rutherford Backscattering 
Spectrometry (RBS), Atomic Force Microscopy (AFM) and Terahertz Time Domain Spectroscopy 
(THz-TDS), are presented.    
\end{abstract}

%\includegraphics{figs/grabs.pdf}

%\item Research highlights item 1
%\item Research highlights item 2
%\item Research highlights item 3

\begin{keywords}
Multi-Strip, Multi-Gap RPCs \sep high counting rate \sep ageing, high irradiation dose \sep SEM, XPS, foil-ERDA, RBS, AFM, THz-TDS
\end{keywords}

\maketitle
\doublespacing
\section{Introduction}

   As it is well known, multi-differential analysis in hadron collisions at
   relativistic and ultra-relativistic energies, especially for rare probes, mandatory for 
   understanding the underlying physics behind the observed phenomena, requires
   unprecedented high statistics. This could be achieved in a reasonable 
   time using high luminosity or high intensity beams and performant 
   experimental setups in high counting rate
   environment. One such example is the CBM
   experiment at FAIR-Darmstadt \cite{cbm1} which is foreseen to cope with 
   interaction rates up to 10$^7$ events/sec for Au-Au collisions at 
   $\sqrt{s_{NN}}$= 2-5 GeV which, due to kinematical focussing, produce 
   charged particles counting rate of $\approx 3$ $\cdot$ 10$^4$ (charged 
   particles)/cm$^2$$\cdot$sec and $\approx$1 hit/$cm^2$ 
   at the smallest polar angles covered by the experiment. One of the main 
   component of the CBM experimental setup is the ToF sub-detector based on multi-gap 
   resistive
   plate counters (MGRPC) \cite{ToF}. 
   Detailed R\&D studies showed that Multi-Strip Multi-Gap Resistive Plate Counters (MSMGRPC) 
   based on float glass
   could reach efficiency better than 95\%, time resolution around 60 psec, position resolution 
   across the strips of $\sigma_y\leq$1.7 mm and along the strips of $\sigma_z\leq$1.55 cm 
   \cite{Pet1, Pet2, Schu}.  FOPI \cite{mlad} was the first large experiment using a time-of-flight
   barrel based on MSMGRPC and among the first ones using MRPCs, HADES \cite{HADES}, STAR
   \cite{STAR} and ALICE \cite{ALICE}.
   The R\&D activity continued in order to develop MSMGRPCs which fulfil the above mentioned 
   requirements for the CBM experimental setup in terms of high counting rate \cite{Bar1}, 
   high granularity
   \cite{Pet3}, tunable transmission line impedance to be matched
   to the frontend electronics \cite{Bar2} with efficiency better than 95\% and time 
   resolution around 50 psec \cite{Petri1}, based on low resistivity Pestov glass \cite{Bar1} or
   Chinese glass electrodes \cite{jwang}. 
   High counting rate test done at COSY-Julich using direct proton beam of 2.5 GeV/c
   showed a time resolution better than 70 ps and efficiency higher than 90\% even at $10^5$ protons/$cm^2$$\cdot$sec \cite{Pet3}.
   Test of the counter performance in high counting rate all over the counter area was done at SIS18-GSI-Darmstadt using charged particles produced by Ni beam of 1.7 A$\cdot$GeV on 1mm thick Pb target up to the highest intensity per spill delivered by SIS18 \cite{ToF}. No deterioration of the counter performance in terms of time resolution and efficiency was observed up to $10^4$ particles/$cm^2$$\cdot$sec.
   As CBM experiment is foreseen
   to run 2 months/year for about 10 years, ageing tests in high irradiation 
   dose are mandatory in order to guarantee that such a detector will 
   maintain its performance over the whole lifetime of the experiment. 
    Timing MRPCs reach very good 
    time resolution and efficiency performance based on multi-gap structure and floating glass electrodes, flushed with C$_2$H$_2$F$_4$ + SF$_6$ + C$_4$H$_{10}$ gas mixture with different relative weights. Therefore, Si from the resistive glass electrodes, known as being highly volatile, with high probability to produce polymeric structure as well as the polymerisation of hydrocarbons from the gas mixture are the main components which contribute to the production of reactive species in polymerisation or chemical deposition in high density avalanches induced by high irradiation dose environment.
    Detailed studies of ageing phenomena, as the result of two competitive phenomena, i.e. ablation and polymerisation,  in low pressure plasma were published in Ref. \cite{HY}. Long term ageing studies for timing MGRPC based on resistive electrodes, at a modest counting rate and total accumulated charge, and analysis of the deposited material on the electrodes were also published \cite{SG}.
    Ageing effects on the performance of the MSMGRPC based on low resistivity glass electrodes in 
    very high irradiation dose and short time exposure, together with detailed analysis of the 
    ablation/etching, deposited material and electric properties of the glass electrodes 
    after the irradiation are reported in the present paper.        
    
   A short 
   description of the detector architecture and test results before the 
   irradiation are presented in Chapter 2. Irradiation experimental setup, its 
   operation, control and measurement results during the irradiation are detailed
   in Chapter 3. In Chapter 4 are presented and discussed the tests after the
   irradiation in terms of dark current and dark counting rate, their temperature dependence and long time behaviour. The irradiated counter efficiency and cluster size were also measured. Visual inspection of different RPC components, the results of
   SEM, XPS, foil-ERDA, RBS, AFM, THz-TDS analysis and resistivity measurements are 
   presented in Chapter 5, 6, 7, 8, 9, 10, 11 and 12, respectively.
   Chapter 13 is dedicated to the conclusions.

\section{MSMGRPC architecture and laboratory tests before irradiation}   

\begin{figure}[h]
	\centering
		\includegraphics[scale=0.2]{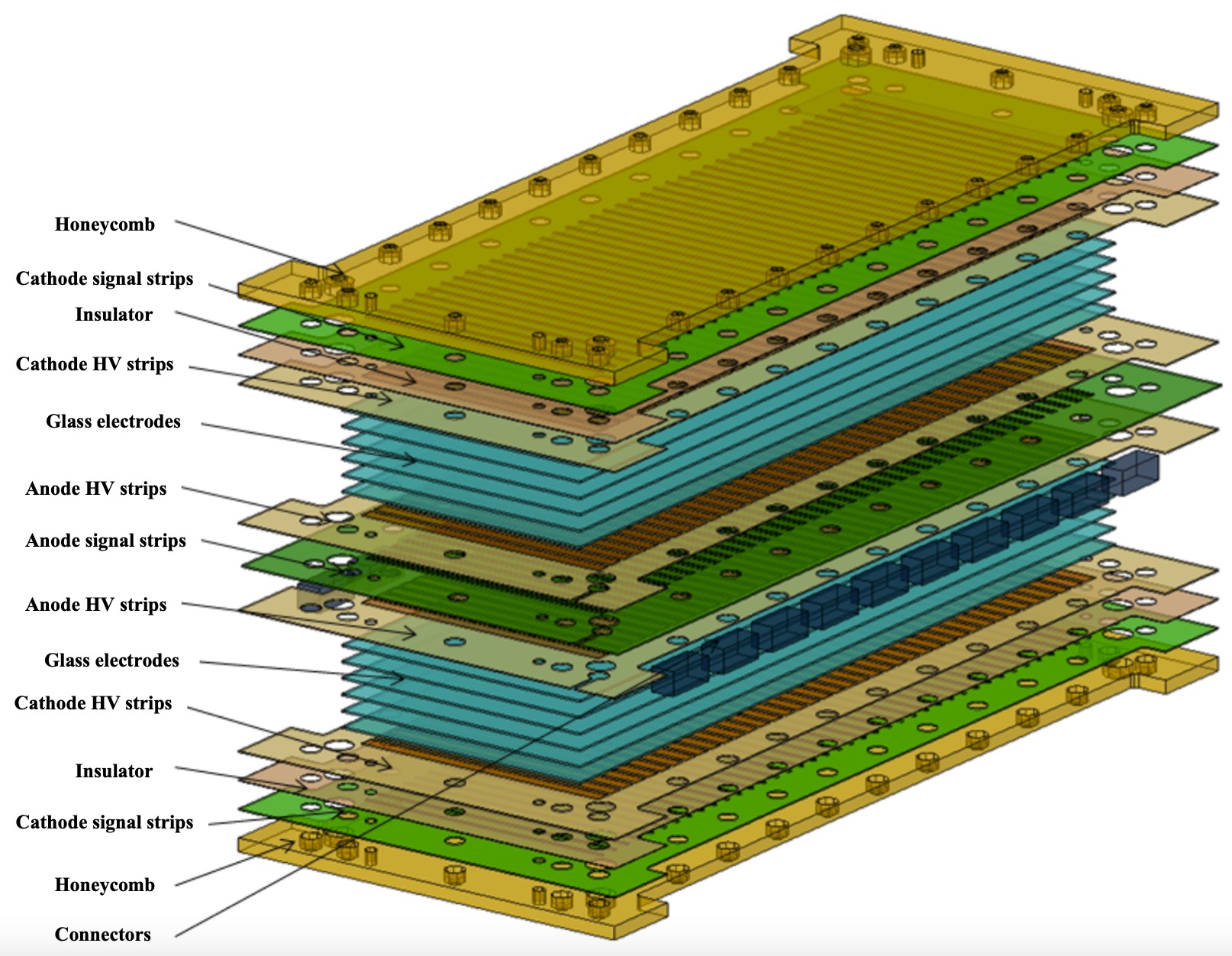}
	\caption{Schematic architecture of a double sided MSMGRPC.}
	\label{FIG:1}
\end{figure}

   The architecture of the double sided (DS) MSMGRPC under investigation is schematically 
   presented in Fig.\ref{FIG:1}. The two halves of the detector are identical and symmetric 
   relative to the central signal strips electrode.   
   Each stack contains six plan parallel resistive electrodes of 0.7 mm thickness, 
   equally 
   spaced by five gas gaps. The size of the gas gaps is defined by the 140 $\mu$m diameter 
   nylon
   fishing line used as spacer. The resistive electrodes are made from low resistivity glass 
   ($\approx$1.5$\cdot 10^{10}$ $\Omega\cdot cm$). The outermost glass plates of each halve are in contact with the Cu strips 
   of the cathode electrodes while the central 2 glass electrode are in contact with the Cu 
   strips of the anode electrode. 
   The strips of the HV electrodes have 5.6 mm width and a pitch size of 7.2 mm. The pick-up 
   signal strips have a width of 1.3 mm with a pitch size of 7.2 mm, as for the 
   HV electrodes. 

 The readout strips behind the corresponding anode and cathode HV ones define a signal 
 transmission line of which impedance depends on the strip width and the properties of the 
 whole structure in between. The signals are readout in a differential mode, both, the 
 anode and the cathode signals, being fed into the input of a readout electronics
 channel, as can be found in Ref. \cite{Bar2}.
   For efficiency and time resolution measurements using cosmic rays, a single sided 
   (SS) MSMGRPC
was assembled. The glass electrodes, HV electrodes, signal read-out printed circuits are 
identical with the ones described above. The counter has 10 glass plates and 8 gaps 
of 140 $\mu$m each,
arranged in 
a single stack. The HV electrodes being only on top and bottom of the structure, 
in order to 
reach the same performance in term of efficiency and time resolution,
the operating voltage is 2 x 9 kV relative to 2 x 5.5 kV for the double sided 
configuration.
This two counters,
mounted in a light electromagnetic screened box, were successfully tested in house 
using cosmic rays and in-beam at
   CERN-SPS with ionising particles produced by a 30 $A\cdot$ GeV Pb beam on Pb target.
   A 62$\pm$3 psec system time resolution and efficiency of $\approx$ 97\% have been
   obtained \cite{Petri2}.    
The final, mandatory test before starting building an 
   important zone of the CBM ToF, i.e. 15$m^2$ in the region of the most forward polar
   angles, is the behaviour of such detectors in high irradiation dose.  
   Simulations based on UrQMD model and FLUKA and GEANT3 transport codes \cite{Anna}
   show that for an Au beam with kinetic energy of 10 A$\cdot$GeV and an intensity of 
   10$^8$ ions/sec on fixed Au target over 2 months period, a ionising dose of
   $\sim$ 500 Gy and a non-ionising dose of $\sim$3$\cdot$10$^{11}$ $n_{eq}/cm^2$ is 
   accumulated in the most inner zone of the CBM ToF. Therefore, we started
   ageing tests of the MSMGRPC described above, using the multipurpose 
   irradiation facility (IRASM) of IFIN-HH \cite{irasm} where a gamma and X-ray doze 
   rate of 0.3 kGy/h, uniform distributed in the area of exposure, can be accessed. 
   
\begin{figure}[h]
	\centering
		\includegraphics[scale=.4]{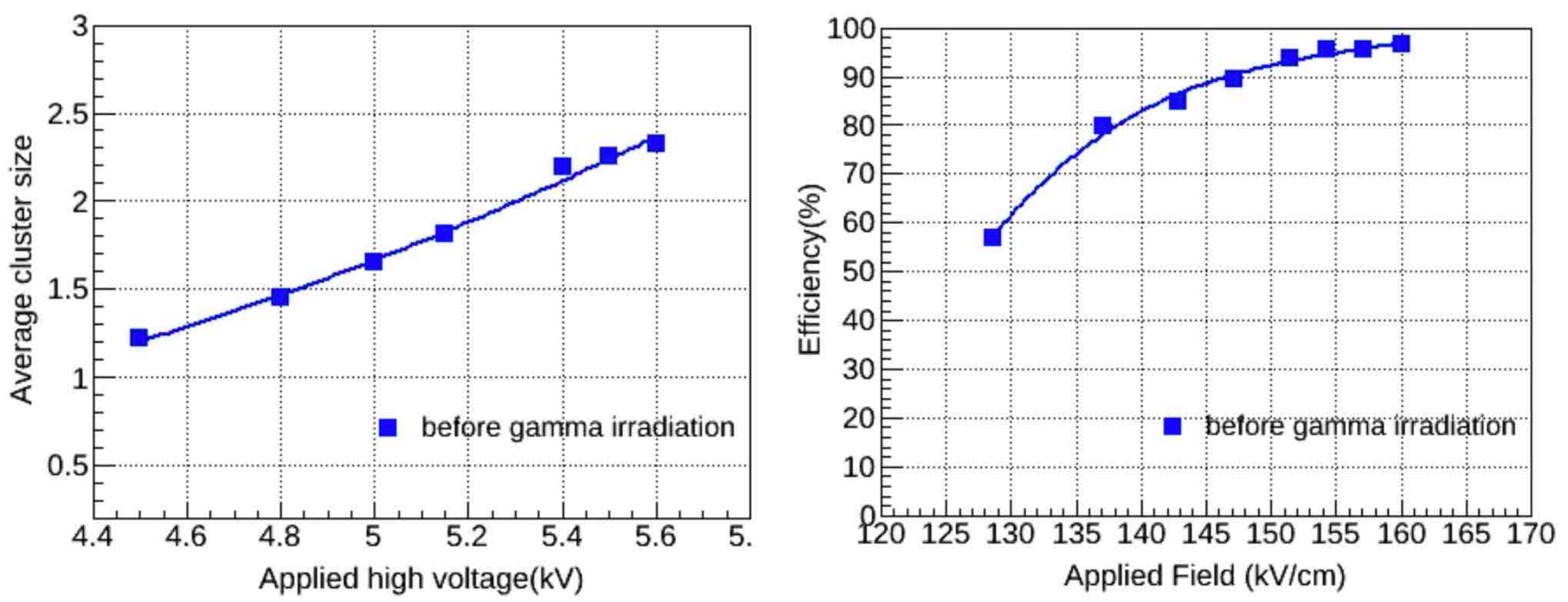}
	\caption{Efficiency (left) and cluster size (right) for cosmic rays tests before irradiation.}
	\label{FIG:2}
\end{figure}

   Before starting the
   irradiation, the DS MSMGRPC counter was tested in the detector laboratory of Hadron Physics 
   Department (HPD) of IFIN-HH with cosmic rays. The counter was disassembled, the glass were 
   carefully cleaned and assembled back, the spacers being arranged along the strips in order 
   to enhance the gas 
   exchange rate through the gaps.   
   16 strips of each of the two RPCs, SS (8 gas gaps, 140 $\mu$m size each) and DS 
   (2x5 gas gaps, 140 $\mu$m size each), were equipped with fast amplifiers based on NINO chip \cite{anghi} and 
   V1290A CAEN TDCs. The housing box of
   $\approx$ 13.5l was flushed with 
   90\%C$_2$H$_2$F$_4$ + 5\%SF$_6$ + 5\% C$_4$H$_{10}$ gas mixture with a rate of 4 liters/hour. Two plastic 
   scintillator bars of 1.5cm x 1.5cm x 10cm where used as trigger in different 
   experimental configurations.
   The results in terms of efficiency 
   and cluster size are presented in Fig.\ref{FIG:2}. The values as a function of the
   applied high voltage obtained in the previous in-house tests are reproduced. At the 
   working voltage, i.e. 5.6 kV, the measured dark
   current was bellow 1 nA while the dark counting rate $\approx$ 1-2 Hz/$cm^2$. 
   
   \section{Irradiation cave and experimental setup}
\begin{figure}[h]
	\centering
		\includegraphics[scale=.22]{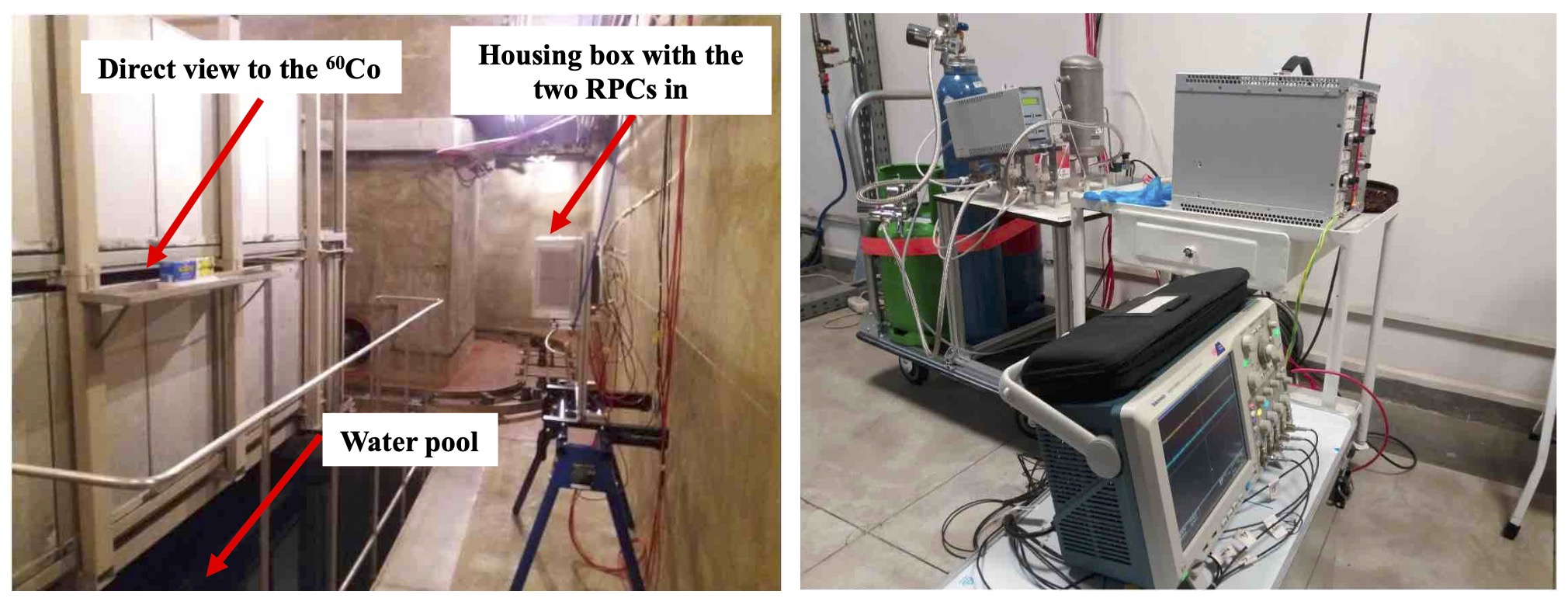}
	\caption{Left photo - irradiation cave, the slit in the middle of the concrete shielding (left) and the
	 MSMGRPC in the housing box (right) equipped with gas transport pipes, HV cables and a signal cable for monitoring the signals directly from the counter; Right photo - the gas mixer, HV units and the oscilloscope placed in a protected area, nearby the
	 control room of the installation.}
	\label{FIG:3}
\end{figure}

   The
   experimental setup was transported and installed in the irradiation cave. In 
   Fig.\ref{FIG:3} left photo can be seen the position of 
   the RPC in the cave. The housing box was aligned (right side, near to the wall) 
   relative to the slit in the concrete protection of the $^{60}$Co source where it is 
   lifted up from the water pool underneath (left side). Long pipes for circulating the
   working gas mixture, HV and signal cables for monitoring the direct signals from the
   RPC were used to connect the RPC to the gas mixer, HV units and oscilloscope placed in
   a protected area, nearby the control room of the installation (right photo of 
   Fig.\ref{FIG:3}).   Once the $^{60}$Co source was lifted in the working position, the
   HV was switched on and slowly increased, carefully monitoring the current and signals 
   on the oscilloscope, mainly to avoid potential micro-discharges. 
 
\begin{figure}[h]
	\centering
		\includegraphics[scale=.45]{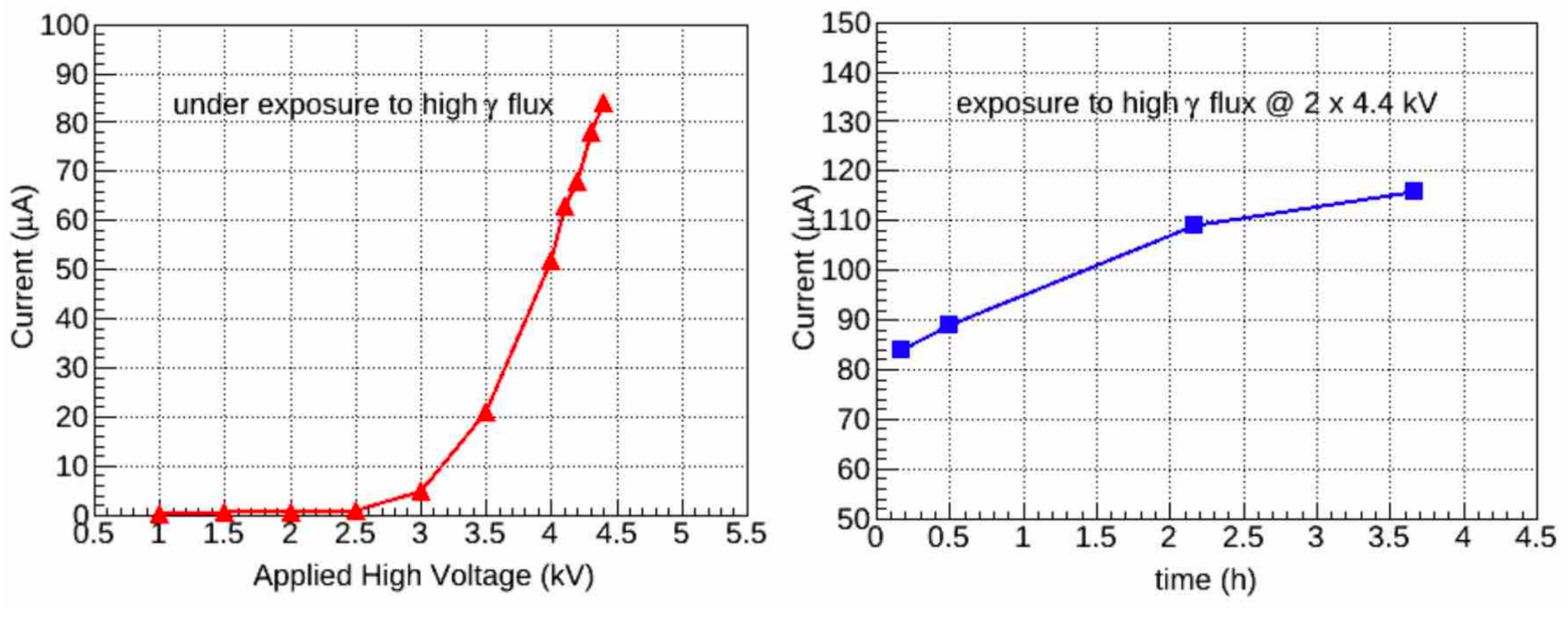}
	\caption{Left - current as a function of applied voltage; Right - evolution of the current with the irradiation time.}
	\label{FIG:4}
\end{figure}

   As expected (see Fig.\ref{FIG:4} left) , above 3.5 kV the current abruptly 
   increases. 
   We decided to stay at 4.5 kV in order to 
   avoid some micro-discharges observed above this voltage. At 4.5 kV the counter efficiency is already in the region of 60\% (see Fig.\ref{FIG:2} right) 
 
 Once we reached this voltage we started to monitor the current as a function of time. 
 Presented in Fig.\ref{FIG:4}-right, the current increases with the exposure
 time. After 2 hours of irradiation, a tendency towards levelling off is observed. These
 plots were obtained during the first irradiation period, a bit longer than 3.5 hours.
  
 The sequence and duration of operating the counter in high irradiation zone, the current, the corresponding charge and accumulated irradiation doze can be followed in Table 1. The total cumulated irradiation doze with HV applied on the counter was 12.99 kGy.
 One has to mention that the total dose during the irradiation slot, including also 
 the periods when the HV was off but the gas flow on, was of 76.7 kGy.   
 
\section{Post irradiation tests}

\begin{figure}[h]
	\centering
		\includegraphics[scale=.42]{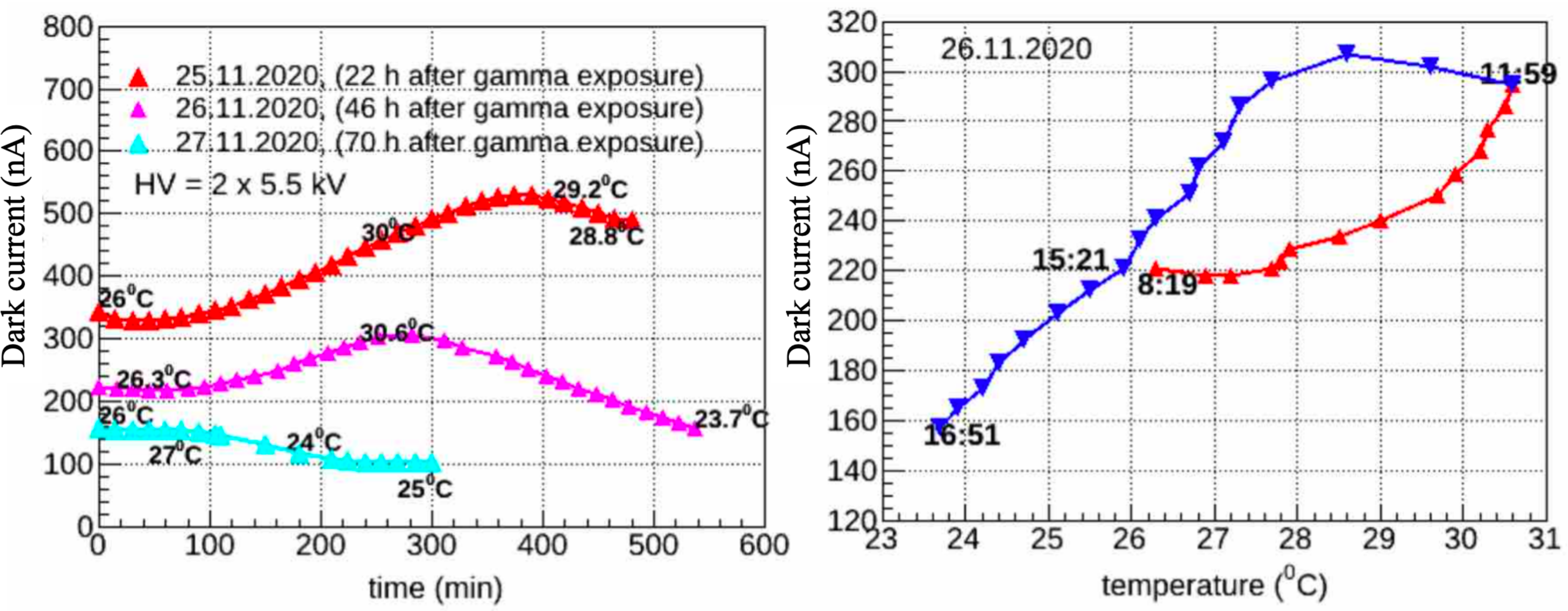}
	\caption{Left - dark current for different periods after irradiation as a function of 
	measured time. The room temperature at different times periods during the measurement could 
	be followed on the three sets of measurements; Right - Evolution of the dark current as a function
	 of room temperature. The time evolution of the temperature during the measurement is marked on the plot at different periods.}
	\label{FIG:5}
\end{figure}

 Once the irradiation period ended and the cave was accessed, the counter and the gas
 mixer were transported back in the detector laboratory of HPD and prepared for detailed 
 tests. The dark current as a function of measured time and room temperature can be
 followed  in Fig.\ref{FIG:5}. One could see that at 22 hours after the irradiation, the
 dark current at 5.5 kV, the value which corresponds to an efficiency better than 95\% 
 is about three orders of magnitude lower than the values measured during the 
 irradiation periods. 

\begin{figure}[h]
	\centering
		\includegraphics[scale=.40]{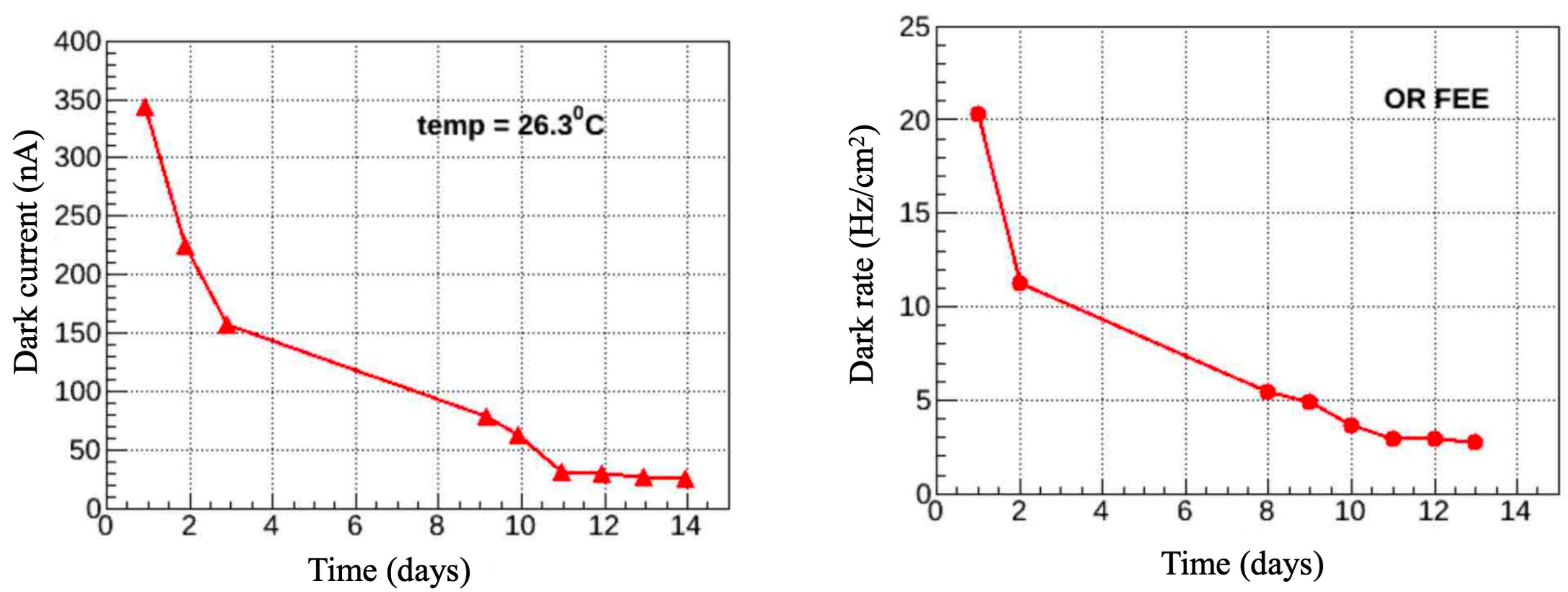}
	\caption{Left - dark current; Right - dark counting rate evolution with time after the irradiation. The counter was operated at nominal voltage.}
	\label{FIG:6}
\end{figure}

It is also evidenced a temperature dependence of the dark current, stronger than the one 
specific for a nonirrardiated counter. The average value and time dependence of the dark 
current decrease with time elapsed since irradiation.
The evolution of the dark current and dark counting rate on a long period is presented in
Fig.\ref{FIG:6}. In two weeks after the irradiation their values converged towards those
measured before the irradiation.
 A similar trend was observed in a long term change of electrical conductance at room temperature in air of a florine graphite fiber intercalation compound
 \cite{TN}.

\begin{figure}[h]
	\centering
		\includegraphics[scale=0.8]{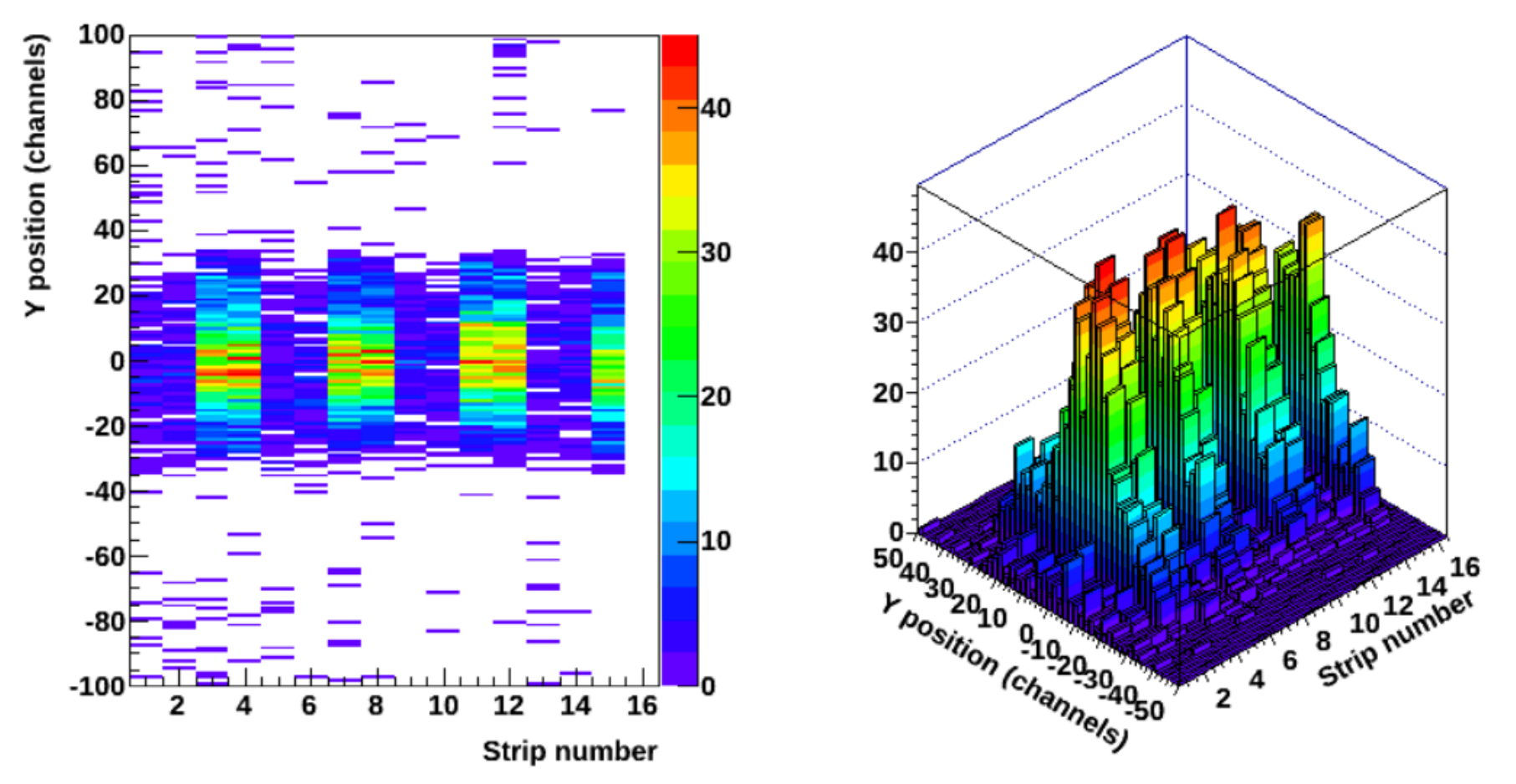}
	\caption{Left - position along the strip versus strip number; Right - the same in a 3-dimensional representation obtained with random coincidence with SS-MGMSRPC.}
	\label{FIG:7}
\end{figure}

Based on the time difference between the two ends
of the 16 operated strips, the position along the strips (y) was 
reconstructed. A two and three-dimensional representations of the position along the strips versus
strip number, obtained as random coincidences with SS-MGMSRPC, is presented in 
Fig.\ref{FIG:7}. 

\begin{table}
\begin{tabular}{|c|c|c|c|c|c|}
\hline
     &        	  &          &         &       &                              \\     
Date & Duration & I       & $<Q>$ & Doze & Cumulated          \\
(dd/mm) & (hours)  &($\mu$A) &   (C) &  rate &  dose        \\  
 &   &. & & (kGy/h)  & (kGy)        \\
 \hline 
     &        	  &          &         &       &                               \\           
10.11& 3:30.    & 105     & 1.4175&  0.3267   & 1.225              \\
 \hline 
     &        	  &          &         &       &                                \\           
11.11& 2:30     & 125     & 1.125&  0.3267    & 2.096               \\
\hline 
     &        	  &          &         &       &                                \\           
12.11& 3:00     & 106     & 1.1448&  0.3267   & 3.076                \\
\hline   
     &        	  &          &         &       &                                \\           
13.11& 3:00     & 168     & 1.8144&  0.3267   & 4.056                 \\
\hline       
     &        	  &          &         &       &                                 \\           
16.11& 3:20     & 289     & 3.468&  0.3222    & 5.130                  \\             
\hline 
     &        	  &          &         &       &                                  \\           
17.11& 3:30     & 363     & 4.5738&  0.3222   & 6.258                  \\
\hline  
     &        	  &          &         &       &                                  \\           
18.11& 6:35     & 254     & 6.0198&  0.3222   & 8.379                  \\
\hline  
     &        	  &          &         &       &                                   \\           
20.11& 4:00     & 397     & 5.7168&  0.3145   & 9.637                  \\
\hline 
     &        	  &          &         &       &                                   \\            
23.11& 3:10     & 233     & 2.6562&  0.3145   & 10.633                \\
\hline 
     &        	  &          &         &       &                                 \\            
23.11& 3:00     & 288     & 3.1104&  0.3145   & 11.577                \\
\hline 
     &        	  &          &         &       &                                   \\            
24.11& 4:30     & 246     & 3.990 &  0.3145   & 12.992                 \\
\hline   
     &        	  &          &         &       &                                   \\            
     & 40:33    &         & 35.0367&          &      12.992            \\ 
\hline           

\end{tabular}
\caption{Date, irradiation duration, MSMGRPC current, total accumulated charge, irradiation dose rate and cumulated irradiation dose during the irradiation tests.}
\end{table}

A clear enhanced activity in the adjacent strips, left-right, to the spacers positioned in 
between them, is evidenced.

\begin{figure}[h]
	\centering
		\includegraphics[scale=.9]{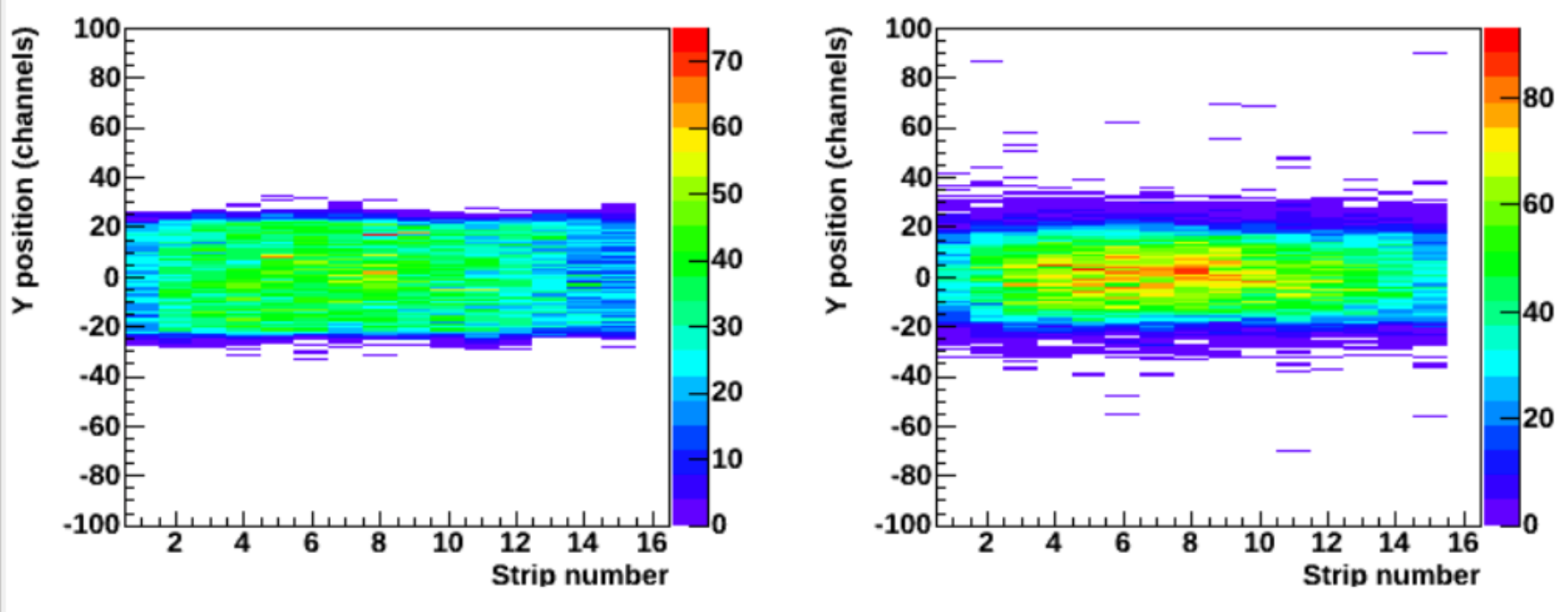}
	\caption{Cosmic ray tests: Left - position along the strip versus strip number for the irradiated double sided MSMGRPC; Right - position along the strip versus strip number for a non-irradiated single sided MSMGRPC used in coincidence with a plastic scintillator as a trigger.}
	\label{FIG:8}
\end{figure}

Using cosmic rays and a coincidence condition between the single sided  MSMGRPC, 
exposed to the irradiation dose without HV on, and one of plastic scintillator bars, such
that the double sided MSMGRPC under investigation to be sandwiched by them,
two-dimensional plots y vs. strip numbers for the two MSMGRPC were obtained. They are 
presented in Fig.\ref{FIG:8}. It is clearly seen that the coincidence condition with the
plastic scintillator removes the activity within the spacers regions and a rather uniform 
distribution of the hits produced by cosmic rays on the surface of the MSMGRPC under 
investigation is obtained.

\begin{figure}[h]
	\centering
		\includegraphics[scale=.55]{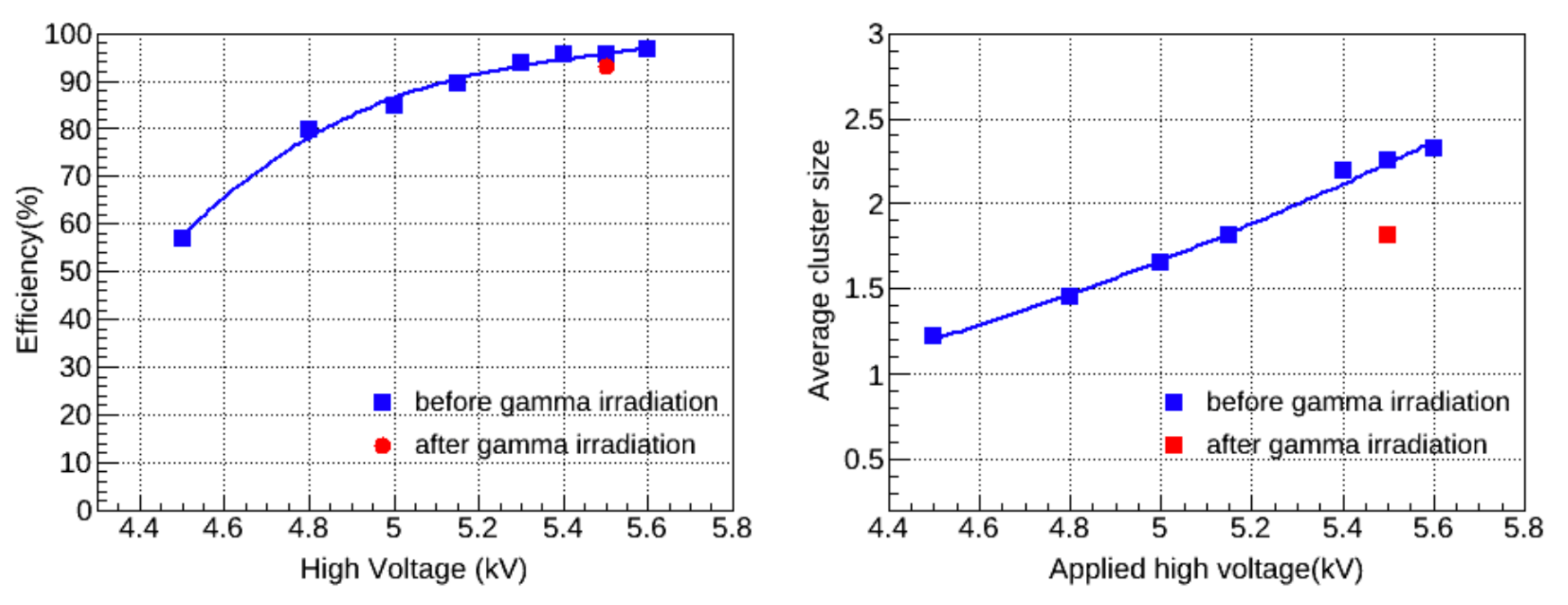}
	\caption{Left - efficiency; Right - cluster size - red squares compared with the values before the irradiation - blue squares.}
	\label{FIG:9}
\end{figure}

   The resulting values of the efficiency and cluster size for the MSMGRPC operated in 
   high irradiation dose are represented by red squares in Fig.\ref{FIG:9}. The blue
   squares represent the corresponding values before the irradiation. Within the symbols
   size, no change in the efficiency after the irradiation is observed. A decrease in 
   the cluster size by 25\% is observed. This could be explained by a decrease of the signals after irradiation, still above the FEE threshold value for the strips corresponding to the centroid of the avalanche but bellow the threshold value for the neighbouring strips.   

\section{Visual inspection after the irradiation}

\begin{figure}[h]
	\centering
		\includegraphics[scale=.42]{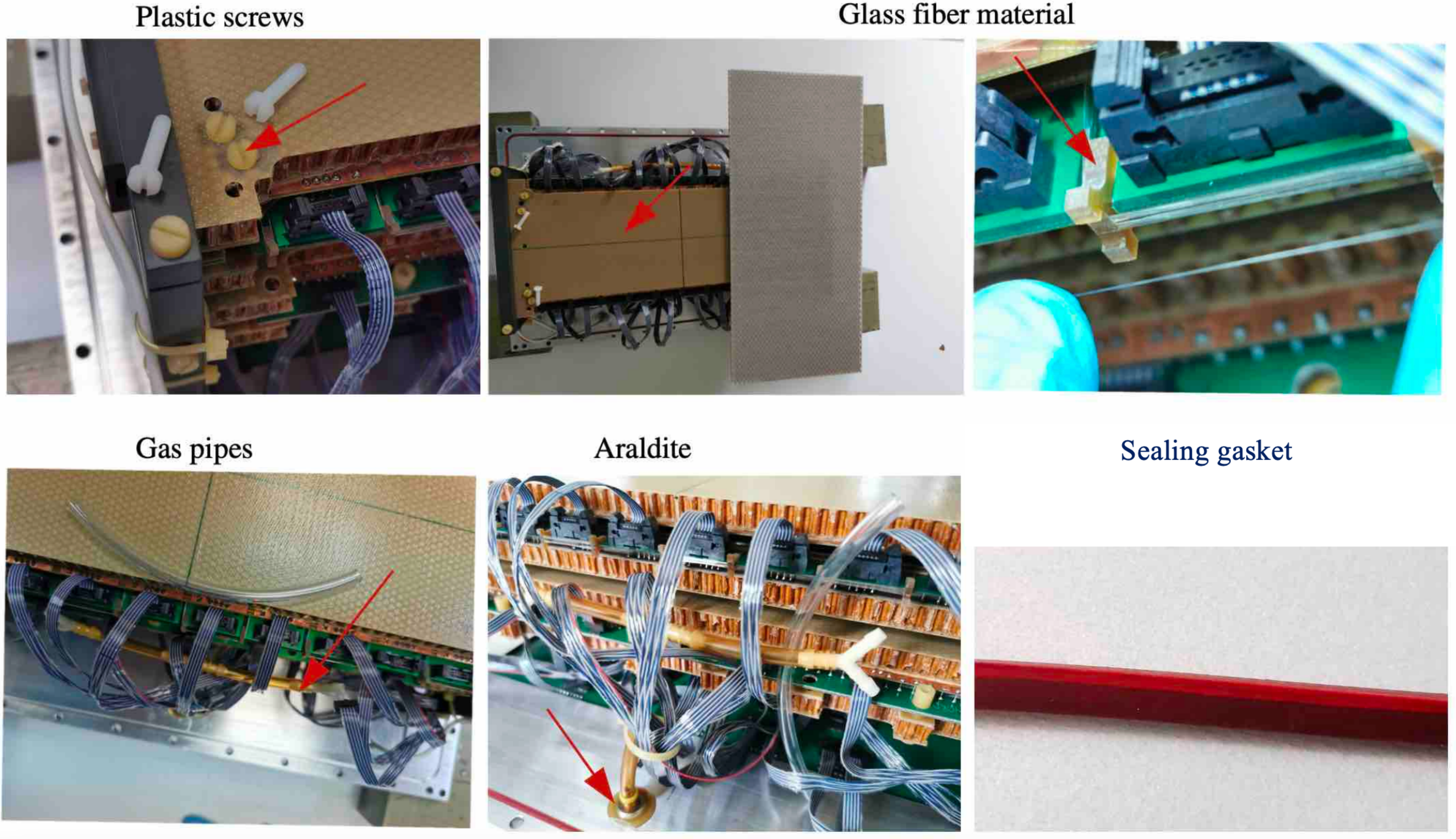}
	\caption{Photos of plastic screws, honeycomb plates used
   to sandwich the counter for mechanical stability, gas pipes compared with non-irradiated
   ones, the epoxy used to seal the HV,
   signals and gas connections through the aluminium back flange and the sealing gasket.}
	\label{FIG:10}
\end{figure}

   After the tests described in the previous chapter, the housing box was opened for a 
   visual inspection of the ageing effects on different detector components. Before 
   disassembling the MSMGRPC, a general inspection showed that plastic screws, the honeycomb 
   plates used
   to sandwich the counter for a mechanical stability,  gas pipes and 
   the epoxy used to seal the HV, signals and gas connections through the aluminium back 
   flange got yellowed, see Fig.\ref{FIG:10}. The gasket used for gas tightness of the housing box became
   rigid. No impact of these changes on the mechanical properties or detector 
   functionality was observed.        
   
\begin{figure}[h]
	\centering
		\includegraphics[scale=.35]{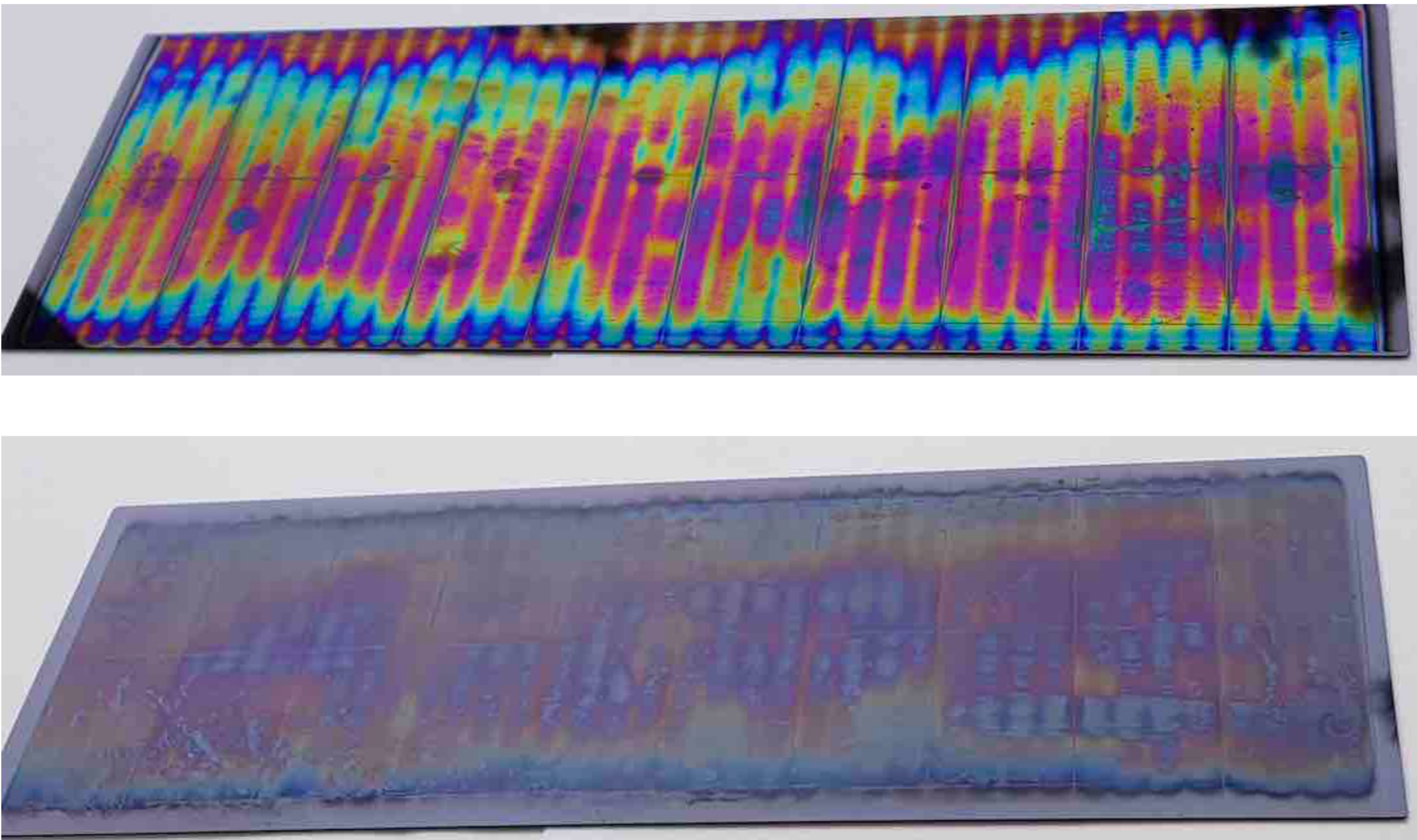}
	\caption{Photos of one of the glass floating electrodes: Top photo - surface facing the cathode (anode): Bottom photo - surface facing the anode (cathode).}
	\label{FIG:11}
\end{figure}

\begin{figure}[h]
	\centering
		\includegraphics[scale=.35]{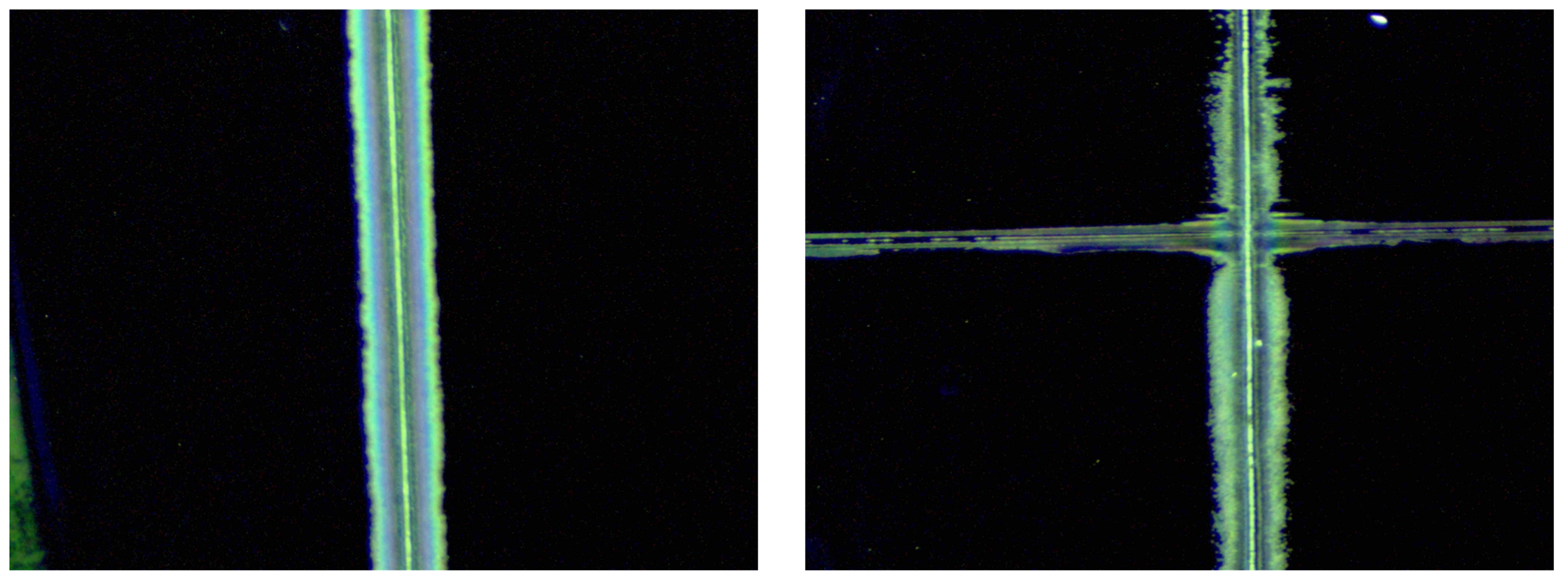}
	\caption{Photos of the regions around the spacers: Left photo - anode surface; Right 
	photo-cathode surface.}
	\label{FIG:12}
\end{figure}

\begin{figure}[h]
	\centering
		\includegraphics[scale=.5]{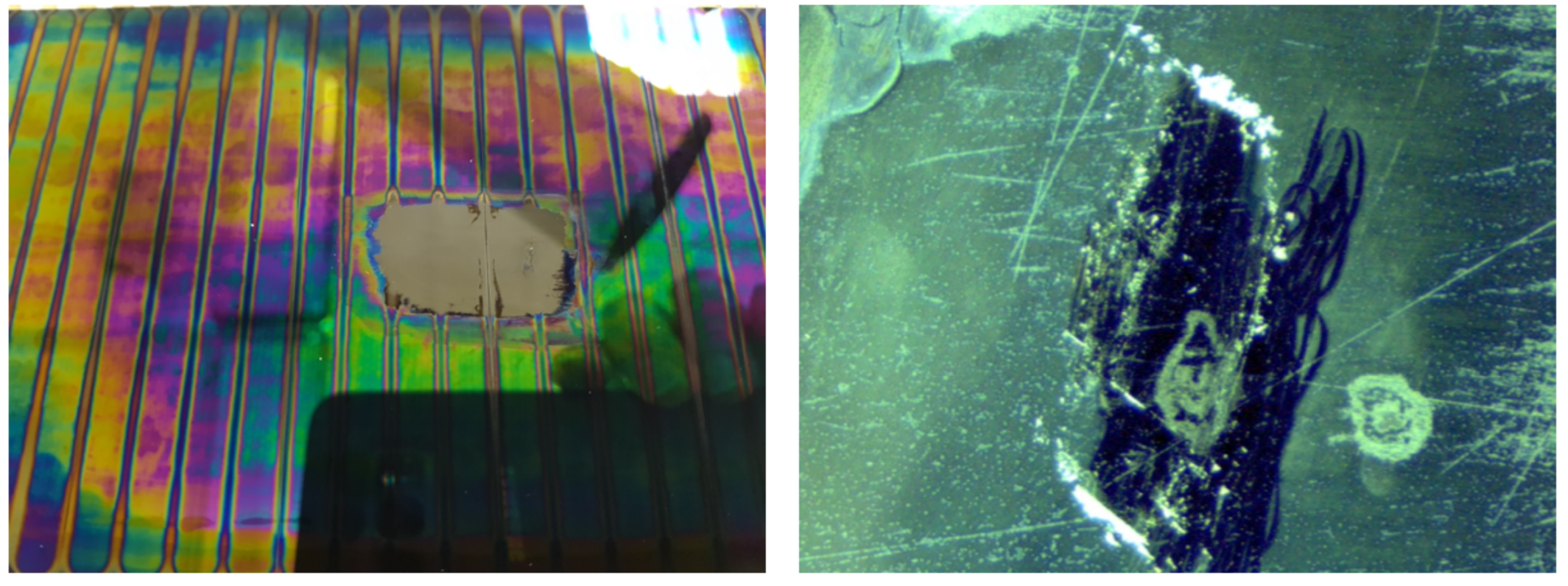}
	\caption{Photos of one of the floating electrodes: Left - cleaned portion of the 
	anode surface using ethyl alcohol; Right - removed deposited layer on the cathode surface 
	using a scalpel.}
	\label{FIG:14}
\end{figure}

   Once the MSMGRPC structure was disassembled we accessed the floating glass electrodes and 
   start the inspection. In the photos from Fig.\ref{FIG:11} could be seen the two sides 
   of a glass floating electrode. The surface facing the cathode electrode is presented 
   in the upper photo and the surface facing the anode electrode in the bottom one. 
   For simplicity, for the rest of the paper we used the label anode for the surface of the 
   glass phasing the cathode and cathode for the surface of the glass phasing the anode. 
   A clear difference, in terms of colour and quantity of the deposited layers, is observed.
   The more consistent and colourful deposited layer seems to be on the anode surface. The 
   strip type pattern corresponds to the strip structure of
   the two high voltage electrodes. All floating glass electrodes present similar 
   characteristics independent on their position within the stack. One can also observe 
   less deposition at
   the edges of the glass electrode where a better gas exchange via diffusion 
   in the gaps region takes place. A closer view of the two surfaces taken with a normal 
   microscope evidenced details of the deposited layers. In   
   Fig.\ref{FIG:12} are shown regions around the spacers.  
   The left photo corresponds to the anode surface and the right
   photo corresponds to the cathode surface. It is observed that in 
   the region of spacers, where the highest dark counting rate was observed, the deposition is
   enhanced. Unexpected thiner horizontal trace in the right photo, cathode surface, is
   evidenced. 
   \begin{figure}[t]
	\centering
		\includegraphics[scale=0.3]{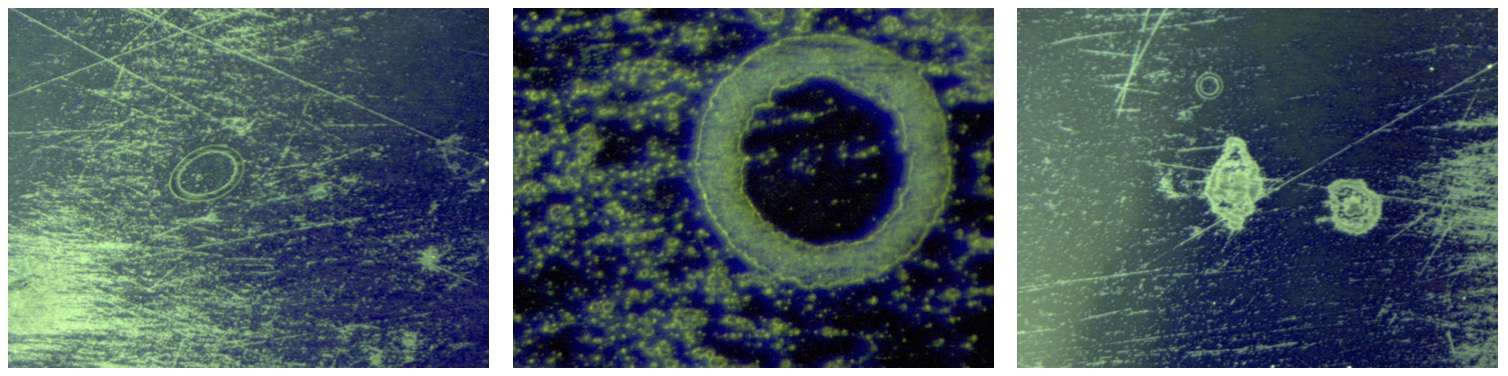}
	\caption{Photos of different regions of the cathode surface of glass electrode.}
	\label{FIG:13}
  \end{figure}
   It corresponds to 
   the position of the previous spacer, orthogonal to the high voltage strips, used in the
   tests before ageing studies. For a better gas exchange in the gas gaps region we decided to
    position the spacers along the HV strips, in the gap region between two adjacent strips. 
    The counter was disassembled, the glass plates were carefully cleaned , no trace along the
     previous longitudinal spacers being observed after. The counter was reassembled for ageing tests, using 
    the same glass plates after being cleaned. In Fig.\ref{FIG:13} are 
    presented photos taken in different regions of the cathode surface of the glass electrode.
    Besides irregular deposition
   spots, regular straight thin lines or ring shape patterns are evidenced. 
   Quite probable this is the result of hydrogen fluoride radical, produced 
   in a polymerisation process in highly dense streamers environment, which develops latent
   very fine patterns \cite{HLM}, already existing on the glass plates before irradiation.  
   While the
   deposited layer on the anode surface can be easily removed using ethyl alcohol,
   the clean spot in the left photo of Fig.\ref{FIG:14}, the deposited layer on the cathode 
   surface of the glass electrode can not be removed using distilled water, 
   ethyl alcohol or acetone. The continuous layer can be removed mechanically
   using a fine scalpel, resulting fine white powder, see the right photo of 
   Fig.\ref{FIG:14}.          
\noindent
However, in the cleaned area remains a spot which seems to be a kind
   of crater in the glass electrode, quite probable the result of an ablation process. After this visual inspections we decided to use 
   different analysis methods in order to find the elemental and chemical composition of
   the deposited layers, their roughness and electrical properties of the glass electrodes of 
   the irradiated 
   MSMGRPC relative to a non-irradiated glass electrode. The results are presented in the 
   following chapters.
   
\section{SEM analysis}

\begin{figure}[t]
	\centering
		\includegraphics[scale=.53]{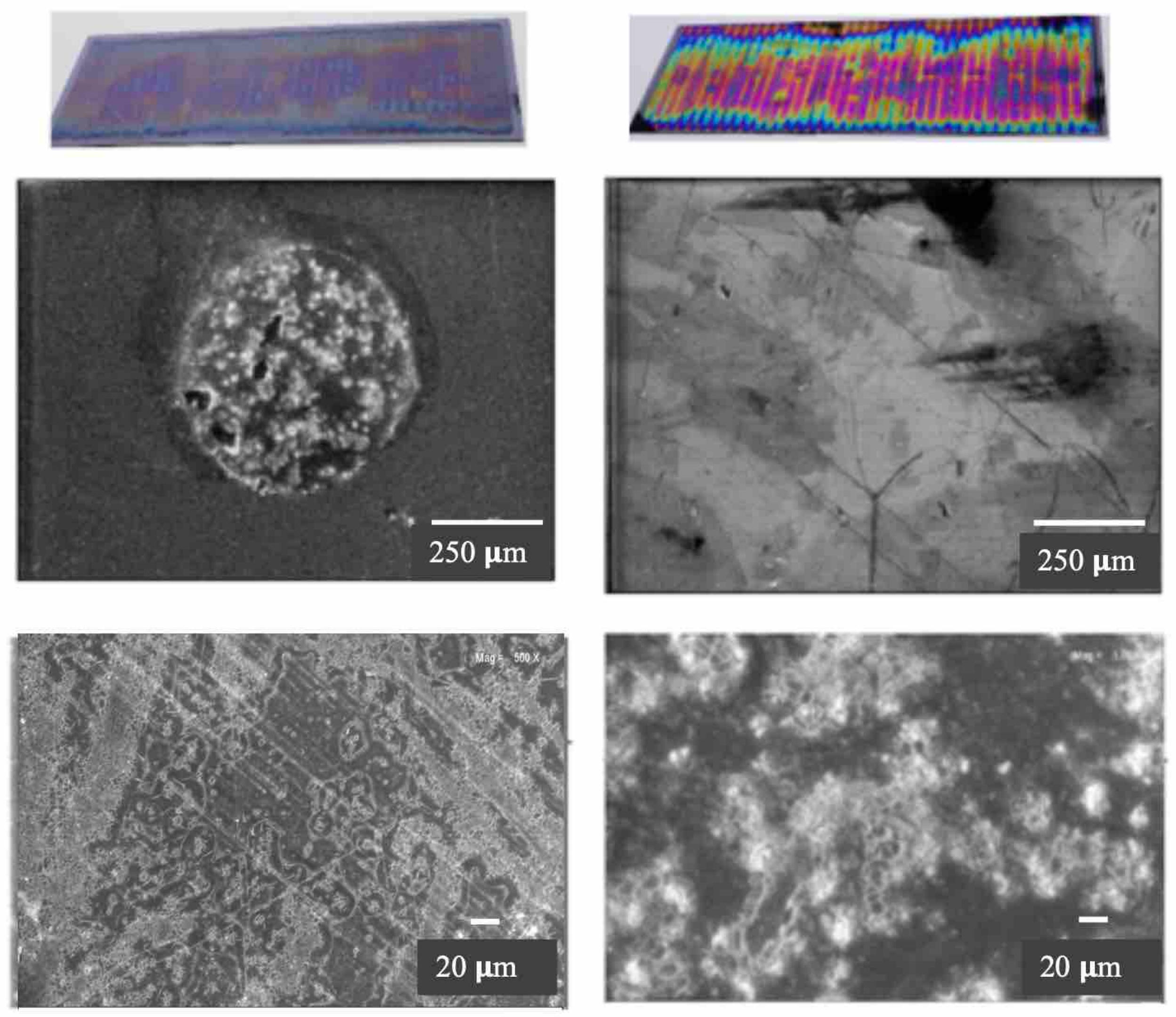}
	\caption{Samples of SEM images of one of the floating electrode surfaces: 
	Left column - the anode surface; Right column - the cathode surface.}
	\label{FIG:15}
\end{figure}

\begin{figure}[h]
	\centering
		\includegraphics[scale=.4]{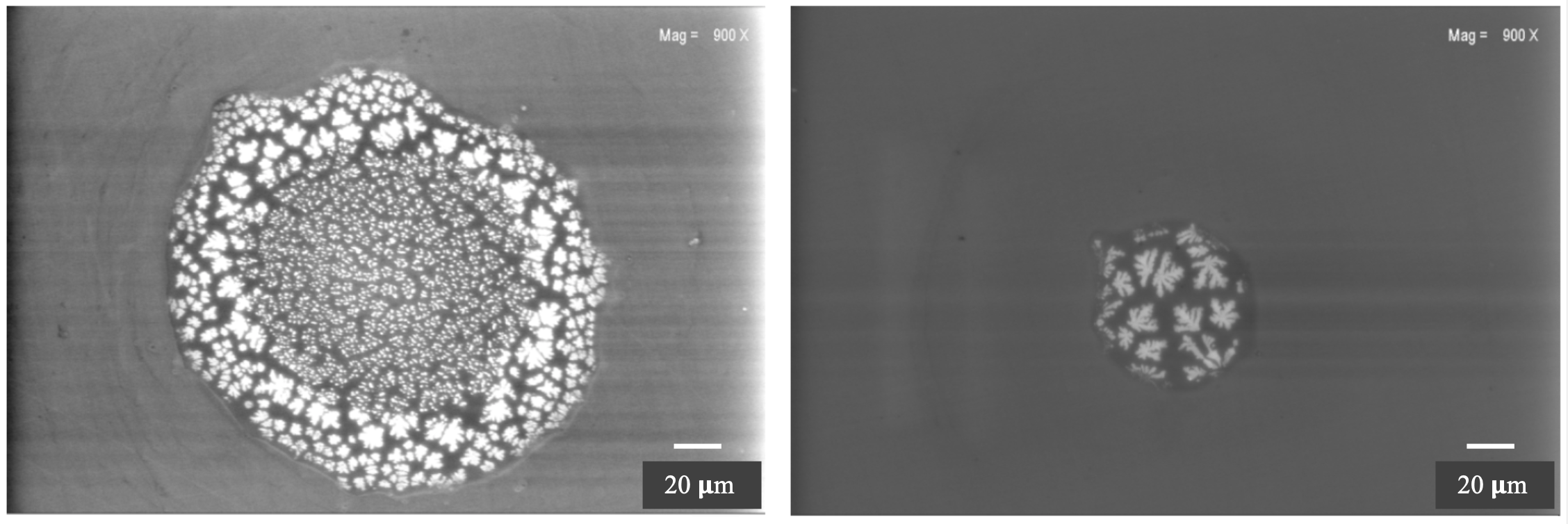}
	\caption{Magnified two regions of a not irradiated glass plate.}
	\label{FIG:16}
\end{figure}

   A scanning of the deposited layer structure with higher resolution was done using 
   a Scanning Electron Microscope (SEM). Samples of some of the obtained images with two
   different resolutions are presented in Fig.\ref{FIG:15}. The left column corresponds to 
   the cathode surface of the glass plate while the right column to the anode surface.
    A clear difference in the structure of deposited layers is evidenced. More regular 
   structures correspond to the cathode surface. A higher resolution image,
   bottom row, evidence irregular spots on the anode surface. Similar
   scanning was done for a non-irradiated glass plate and the results are presented in
   Fig.\ref{FIG:16}. While by naked eye such zones hardly could be spotted, with SEM 
   method one could observe regular patterns, ring or disc type which clearly shows that
   they have a crystalline structure. Such local structures were also observed in
   Pestov type low resistivity glass \cite{pestov} and expected to be produced during the 
   rather complex production process, rather different than the one used for commercial
   float glass production.     
       
\begin{figure}[t]
	\centering
		\includegraphics[scale=.35]{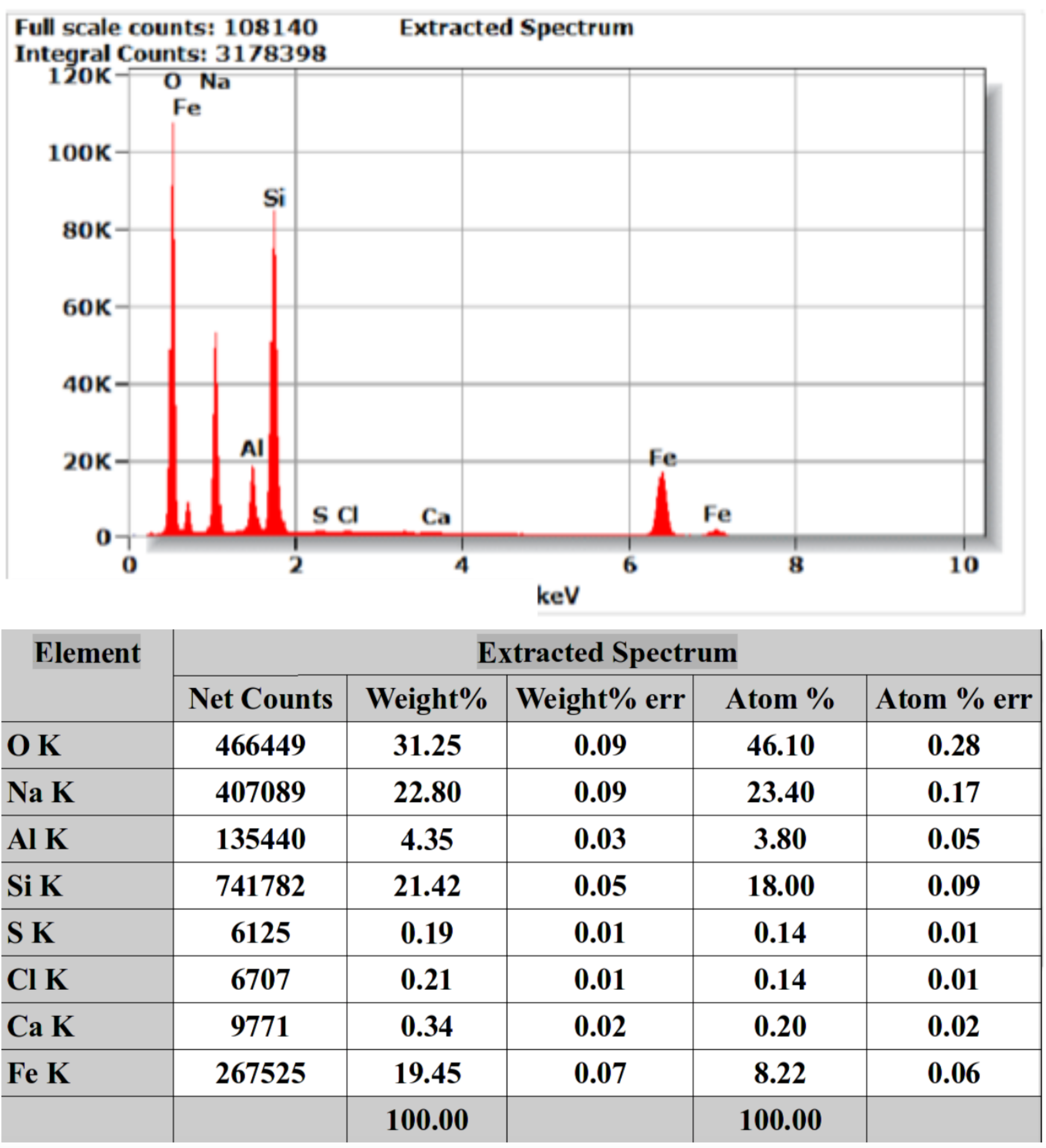}
%		\vspace{-2.5cm}
	\caption{Top: electron spectra corresponding to the surface of the non-irradiated glass plate; bottom: the corresponding relative elemental composition.}
	\label{FIG:17}
\end{figure}

\begin{figure}[t]
	\centering
		\includegraphics[scale=.35]{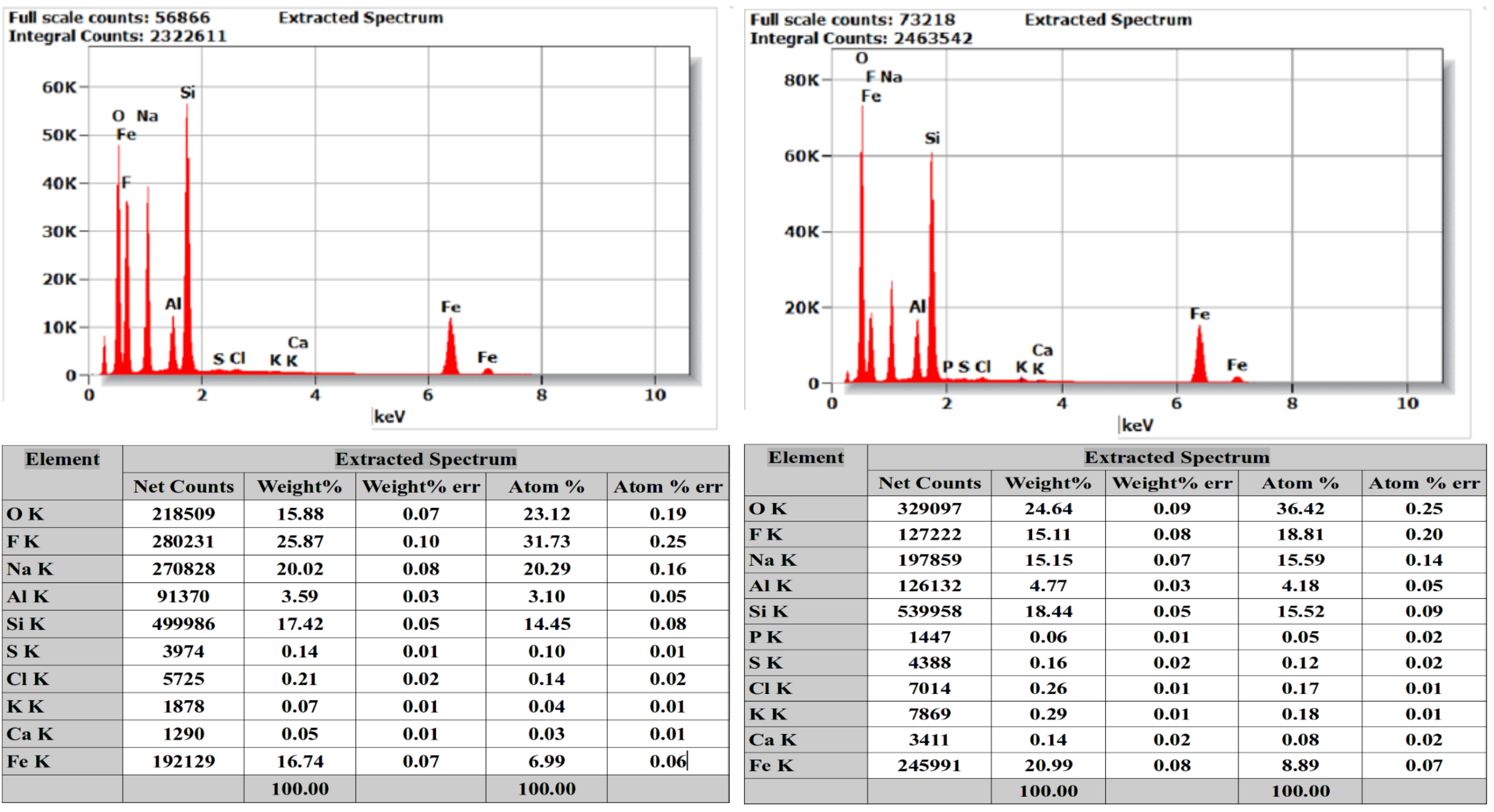}
	\caption{Left-top: electron spectra corresponding to anode surface of irradiated glass;
	Left-bottom: the corresponding relative elemental composition. Right-top: 
	electron spectra corresponding to cathode surface of irradiated glass;
	Right-bottom: the corresponding relative elemental composition.}
	\label{FIG:18}
\end{figure}

 Further Energy-Dispersive X-ray Spectrometry (EDX) investigations on the 
 elemental 
 composition of some samples of surfaces were done and the results are presented in the
  top plots of Fig.\ref{FIG:17}
 and Fig.\ref{FIG:18} for a non irradiated glass and for the anode and cathode surfaces of 
 a glass electrode after irradiation of MSMGRPC, 
 respectively. The EDX analysis was done in standard less mode.
 The relative content of different components are listed in bottom tables of each 
 figure. The elements found by analysing the surface of a not exposed glass plate confirm
 the elemental composition reported by the producers, i.e. SiO$_2$, Fe$_2$O$_3$, Na$_2$O, 
 Al$_2$O$_3$, MnO$_2$ \cite{yw} with the exception of Mn. 
 The lack of manganese from EDX spectra can be explained by the fact that the amount in 
 the samples is below the detection limit of $\approx$ 1,000 ppm.
 As far as concerns the relative 
 percentage presented in Fig.\ref{FIG:17} of the present analysis, they can not 
 be confronted with the producer's values as far as they were not reported in the 
 literature.     
  
%\begin{figure}[h]
%	\centering
%		\includegraphics[scale=.5]{Fig18.pdf}
%	\caption{Top: electron spectra corresponding the cathode surface of irradiated glass; 
%	bottom: the corresponding relative elemental composition.}
%	\label{FIG:18}
%\end{figure}
 \begin{figure}[]
	\centering
		\includegraphics[scale=.3]{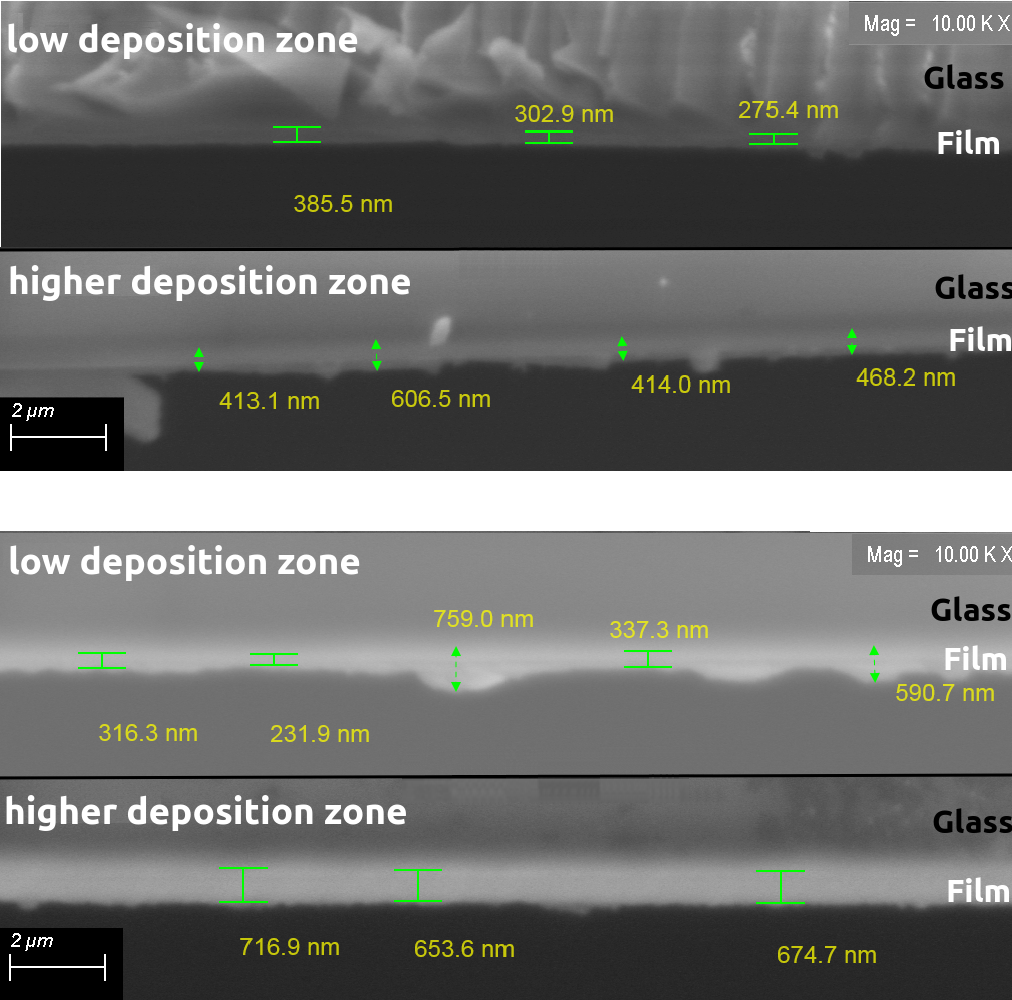}
	\caption{Top: cross-section scanning of the cathode surface in two regions with 
	different deposition thickness, i.e low deposition and higher deposition; Bottom:
	 cross-section scanning of the anode surface.}
	\label{FIG:19}
\end{figure}  

However, a definite conclusion is that the not exposed glass plates do not have 
 any fluorine content. The fluorine percentage on the surface of glass electrodes used
 in the MSMGRPC operated in high irradiation dose is significant and different for the two
 surfaces. The anode surface contains 31.73$\pm$0.3\% fluorine 
 relative 18.81$\pm$0.2\% found on the cathode surface. As far as concerns the 
 oxigen content, relative to 46.10$\pm$0.3\% corresponding to not irradiated glass plate, the
 values obtained for the irradiated glass plate are 23.12$\pm$0.2\% and 36.42$\pm$0.3\% for
 the two surfaces, respectively. For the rest of elements, no significant difference is
 evidence between the irradiated and non-irradiated glass plates. One should
 consider that due to a rather large non-uniformity in the deposited layer on the two surfaces of
 the irradiate glass, the results of the above measurements are position dependent.   
   The analysis was performed in variable pressure secondary electrons (VPSE) mode at 20 Pa and 20 kV
   voltage for accelerating the electron beam. 
   The probe current was 700 pA at a working distance (WD) of 14 mm using different magnifications: x100, x250 and x1,000. A scanned area of $\approx$1.2 $mm^2$ and a penetration depth 
   of 4 $\mu$m were used for the EDX spectra.  
    The results of the SEM investigations performed on glass sample cross section, the glass being cut in several regions and placed at 
    90$^{\circ}$ orientation in the SEM microscope, are presented in 
    Fig.\ref{FIG:19}. An average variation of 50\% and in some particular regions up to a factor of two in the deposited layer is observed. As preliminary optical inspection also suggested, the deposition on the anode surface is considerably thicker than the one deposited on the cathode surface.

\section{XPS analysis}  

\begin{figure}[h]
	\centering
		\includegraphics[scale=.4]{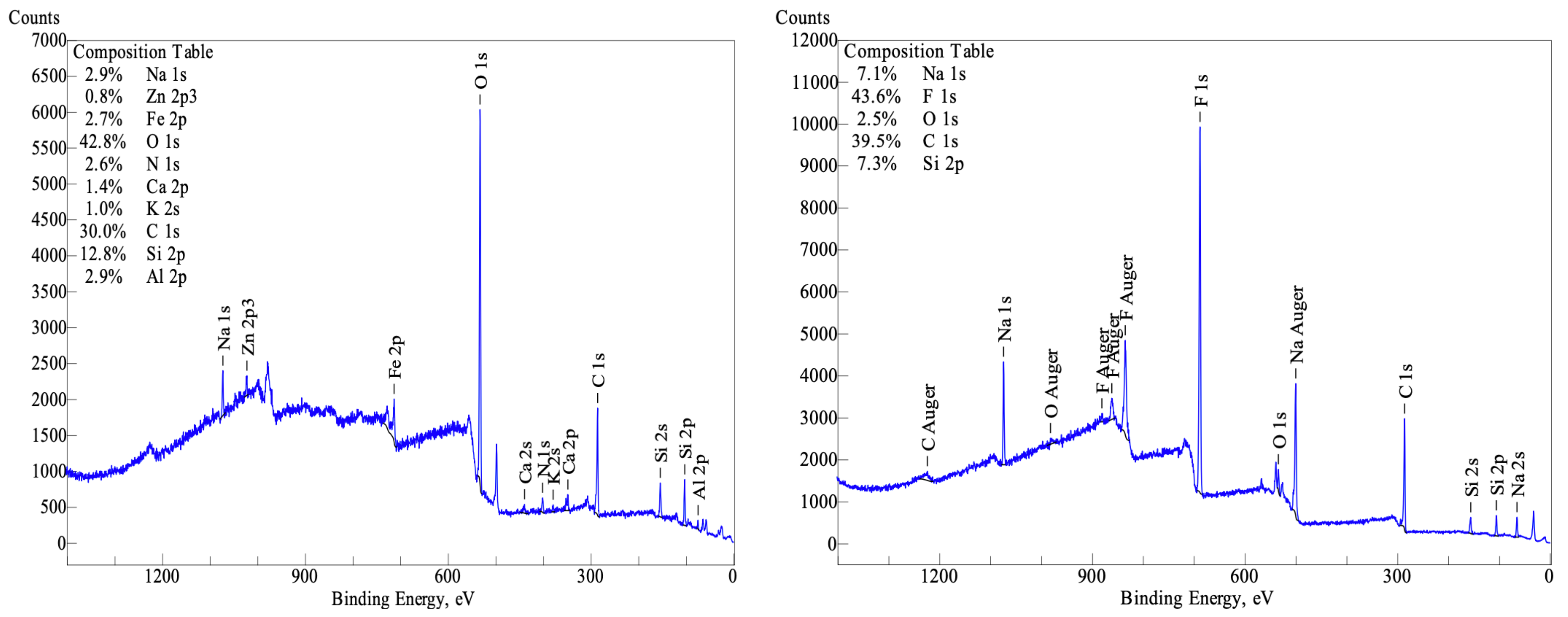}
	\caption{Left: electron spectra for the surface of non-irradiated glass plate; Right: electron spectra for the anode surface of an irradiated glass plate}
	\label{FIG:20}
\end{figure}

 In order to access information on chemical compounds in the deposited layers, we used 
 X-ray photoelectron spectroscopy (XPS) which, based on the photoelectric effect, could give information on elemental composition as well as their chemical state in the
 material. Two samples of electron spectra for the surface of  non-irradiated glass plate and the anode surface of an irradiated glass plate are presented in  
 Fig.\ref{FIG:20} left and right, respectively.  

 Their unfolding gives information on type of chemical bounds of different elements.
 In terms of fluorine composition the results of XPS analysis confirm the findings based on 
 the SEM described in the previous chapter. Darker regions on the anode surface
 have higher fluorine content. As far as concerns the chemical composition, the anode surface of 
 the glass plate used in the irradiated MSMGRPC contains C-F bounds, NaF,
 metallic or Na oxides, CF$_4$ and the cathode surface of the glass plate used in the irradiated 
 MSMGRPC contains C-F bounds, metallic and F oxides. A general conclusion
 is that the metallic oxides are preferentially deposited on the cathode surface 
 while the percentage of fluorine components is higher on the anode surface.  
                      
\section{foil-ERDA analysis}
\begin{figure}[t]
	\centering
		\includegraphics[scale=.2]{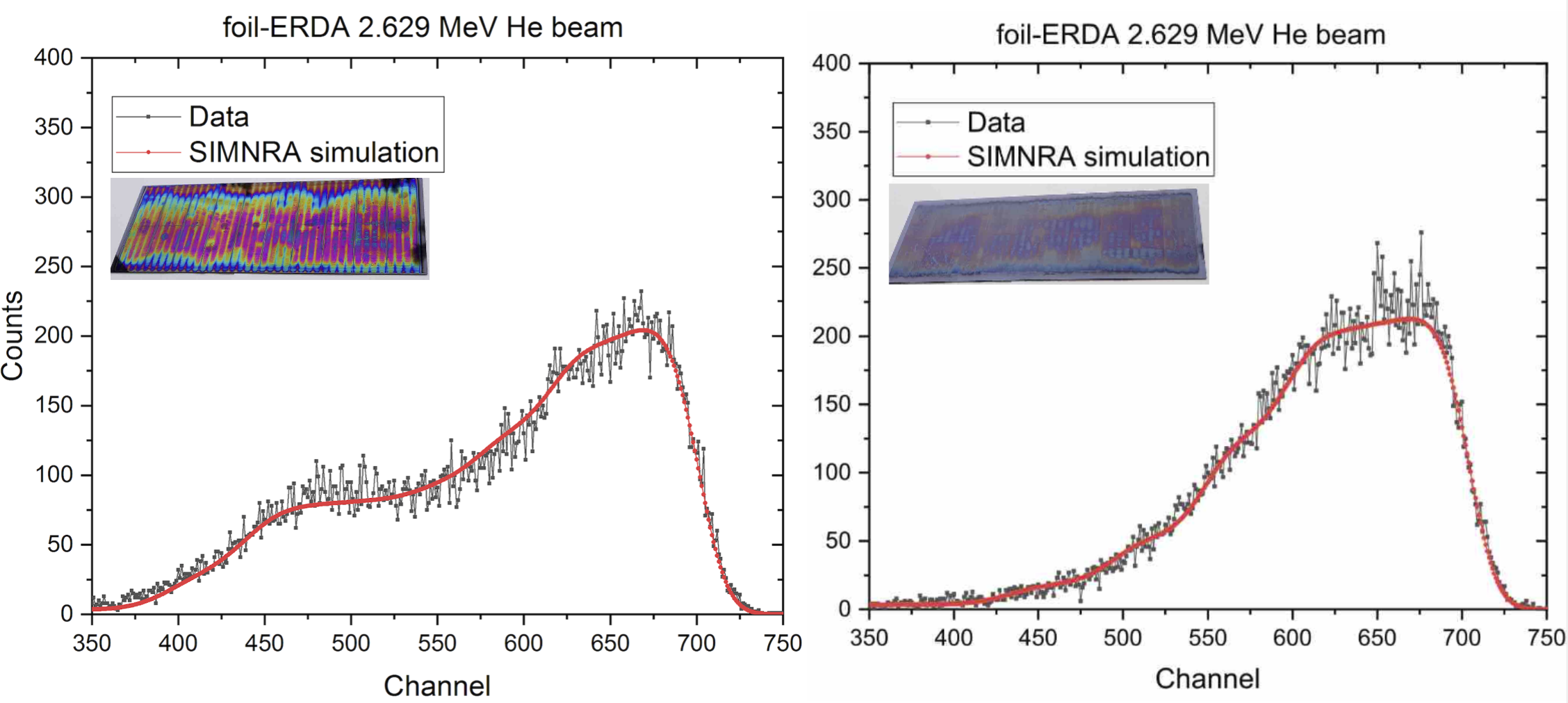}
	\caption{ERDA measurements of hydrogen distribution for the two surfaces of an exposed glass electrodes. ERDA spectra collected with the detector solid angle of 1 msr and the total charge of 1.8 $\mu$C: 
	Left-anode surface; Right-cathode surface. The channel vs. counts spectra can be converted to a depth scale (10$^{15}$ atoms/$cm^2$) vs. H concentration (\%) plot using SIMNRA code. H concentration values and their depth profiles are given in Tables 2 and 3.}
	\label{FIG:21}
\end{figure}

The composition and hydrogen profile of the layers deposited on the glass surface were measured by foil elastic recoil detection analysis (ERDA) and Rutherford backscattering spectrometry (RBS) using ion beams delivered by a 3 MV TandetronTM from IFIN-HH 
 \cite{burd1}. In Fig.\ref{FIG:21} are presented foil-ERDA spectra for the two surfaces of an exposed glass electrode obtained using 2.629 MeV  $^4He^+$ beam.
 Left spectra corresponds to the anode surface while the right corresponds to the cathode surface. The energy of hydrogen recoil was recorded using an AMETEK type BU-012-050-500 solid-state detector with a solid angular acceptance of 1 msr that was placed at a recoiling angle of 
 30$^{\circ}$ relative to the beam. 
The incident 
   beam angle and the exit angle as measured from the normal to the sample surface
    were both 75$^{\circ}$. The 12 $\mu$m thick Mylar$^{TM}$ stopping foil were located in front of
     the ERDA detector to separate hydrogen recoils hitting the ERDA detector from 
     scattered He ions.    
 The RBS and foil-ERDA spectra 
     were simulated using SIMNRA software package version 7.02 \cite{May}. The accumulated charge for 
     each RBS spectrum was 10 $\mu$C while for foil-ERDA was 1.8 $\mu$C. The 
     measurements were performed using a low beam density current, the intensity for
      the electrical current was about 4 nA and the beam spot of around 1.5 mm in 
      diameter.
 
\begin{table}
\begin{tabular}{|c|c|c|c|c|c|c|c|}
\hline
Layer & C	& H	 & F	 & Si & Al	& Na & Thickness\\	
 &                &&&& &&($10^{15}$ atoms/$cm^2$)\\
\hline
 1& 0.43	& 0.18	& 0.282	& 0.05	& 0.01 &	 0.048 & 675 \\
 \hline
2&	0.32&	0.17&	0.297&	0.05	 &0.01&	0.153&	675\\
\hline
3&	0.2&	0.135&	0.407&	0.05& 0.01&	0.198&	555\\
\hline
4&	0.18&	0.085&	0.417&	-&	0.01&0.198&	400\\
\hline           

\end{tabular}
\caption{Relative percentage of elemental content of different layers
	 (first column) and corresponding thickness (last column) for the deposition on the 
	 anode surface of the glass plate.}
\end{table}

\begin{table}
\begin{tabular}{|c|c|c|c|c|c|c|c|c|c|}
\hline
Layer & C	& H	 & F	 & Si & Al	& Na & O & Fe & Thickness\\	
 &                &&&& &&&&($10^{15}$ atoms/$cm^2$)\\
\hline
1& 0.085&	0.19&	0.056&	0.1& 	0.05	& 0.212& 	0.202 & 0.105& 675\\
\hline
2&	0.01&	0.19&	0.056&	0.183&	0.05	& 0.185& 	0.236&	0.09	& 675\\
 \hline
3&	-&	0.13&	0.056&	0.182&	0.05&	0.187&	0.304&	0.09	 & 555\\
\hline
4&	-&	0.06&	0.056&	0.0182&	0.05&	0.188&	0.374&	0.09& 	555\\
\hline
5&	-&	0.026&	0.056&	0.182&	0.05&	0.188&	0.408&	0.09	& 555\\
\hline           

\end{tabular}
\caption{Relative percentage of elemental content of different layers
	 (first column) and corresponding thickness (last column) for the deposition on the 
	 cathode surface of the glass plate.}
\end{table}

 As it is known, this method gives access to a profile of 
 the elemental contribution in different layers in the deposited material. 
 The thickness
 of different layers is given in thin film unit (TFU) and the layers are labeled from 1 to 4(5)
  starting from the surface. The results are presented in Tables 2 and 3.
 The areal density, or thin film units, are the natural units for ion beam analysis, since the
  energy loss is measured in eV/(atoms/cm$^2$) and the monolayer is of the order of 10$^{15}$
 atoms/cm$^2$
 \cite{IBA,Jey}.
 Fluorine is present on both surfaces. The content of fluorine on the cathode surface is lower
  and constant as a function of depth in the deposited layer, relative to the content in the 
  deposited layer on the anode surface where the amount is increasing going deeper in the 
  deposition. Contrary to the anode surface, the cathode one
	contains important percentage of oxygen, increasing with depth. The carbon content is 
	reversed, i.e. the cathode layer has rather negligible amount of carbon in the superficial 
	layer while the percentage of carbon in the anode layer is high, 
	decreasing going deeper in the deposited layer.
 
\section{RBS analysis}

\begin{figure}[h]
	\centering
		\includegraphics[scale=.45]{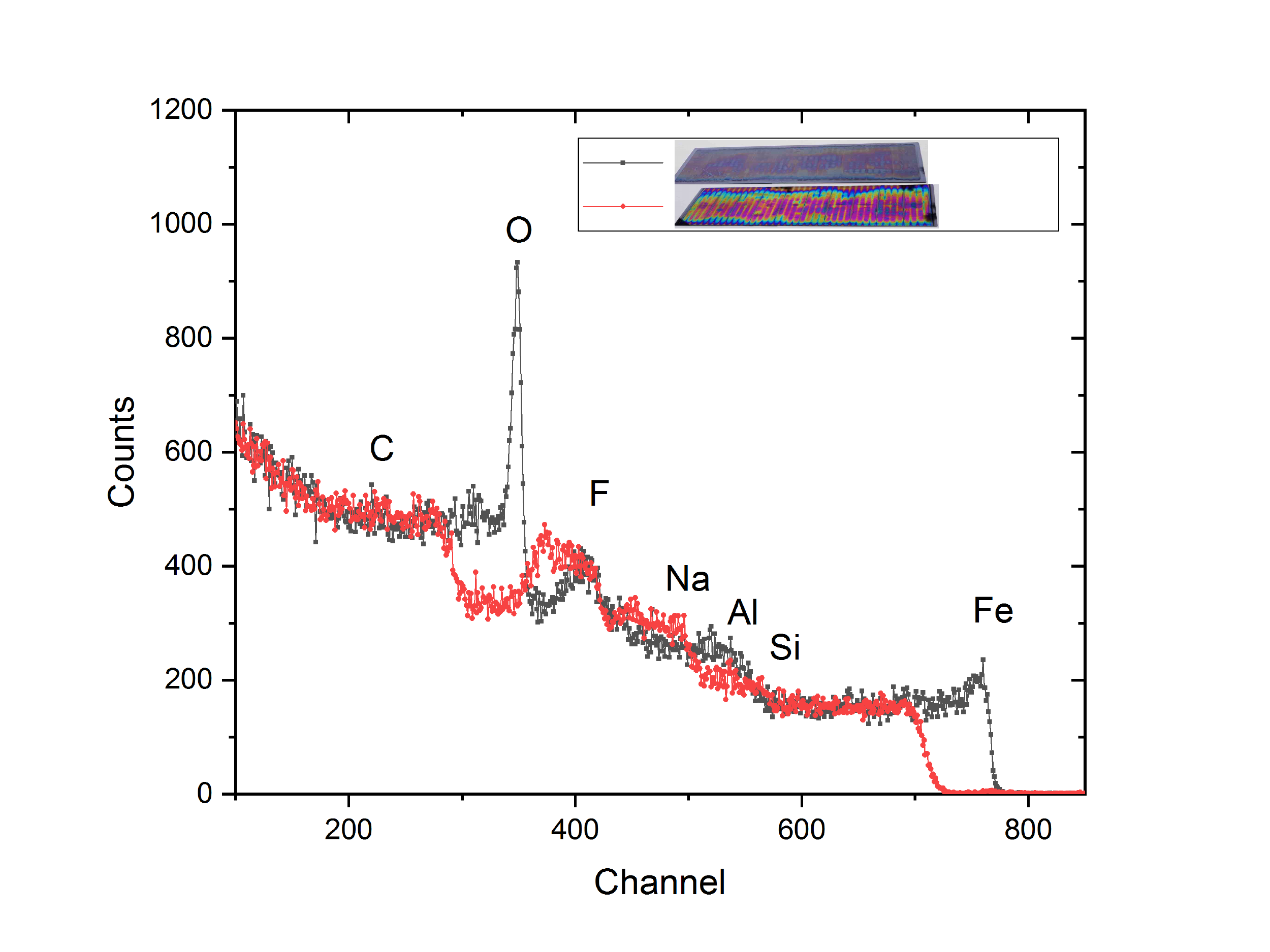}
	\caption{RBS spectra corresponding to the two surfaces of irradiated glass plate overlapped: Black colour corresponds to the cathode surface, Red
	colour corresponds to the anode surface.}
	\label{FIG:22}
\end{figure}

Non-Rutherford Backscattering Spectrometry (NRBS) using 3.04 MeV and 4.28 MeV He beams were performed to determine the thickness and stoichiometry of the samples. A Si detector was placed at 165$^{\circ}$ with respect to the beam direction. The energy resolution of the experimental setup was about 18 keV. In order to quantify the presence of light elements (C and O) in the sample we have used 3.04 MeV to take advantage of the high cross-section of $^4$He with oxygen and 4.28 MeV with carbon.
	The results are presented in Fig.\ref{FIG:22}. 
\noindent

\begin{figure}[h]
	\centering
		\includegraphics[scale=.35]{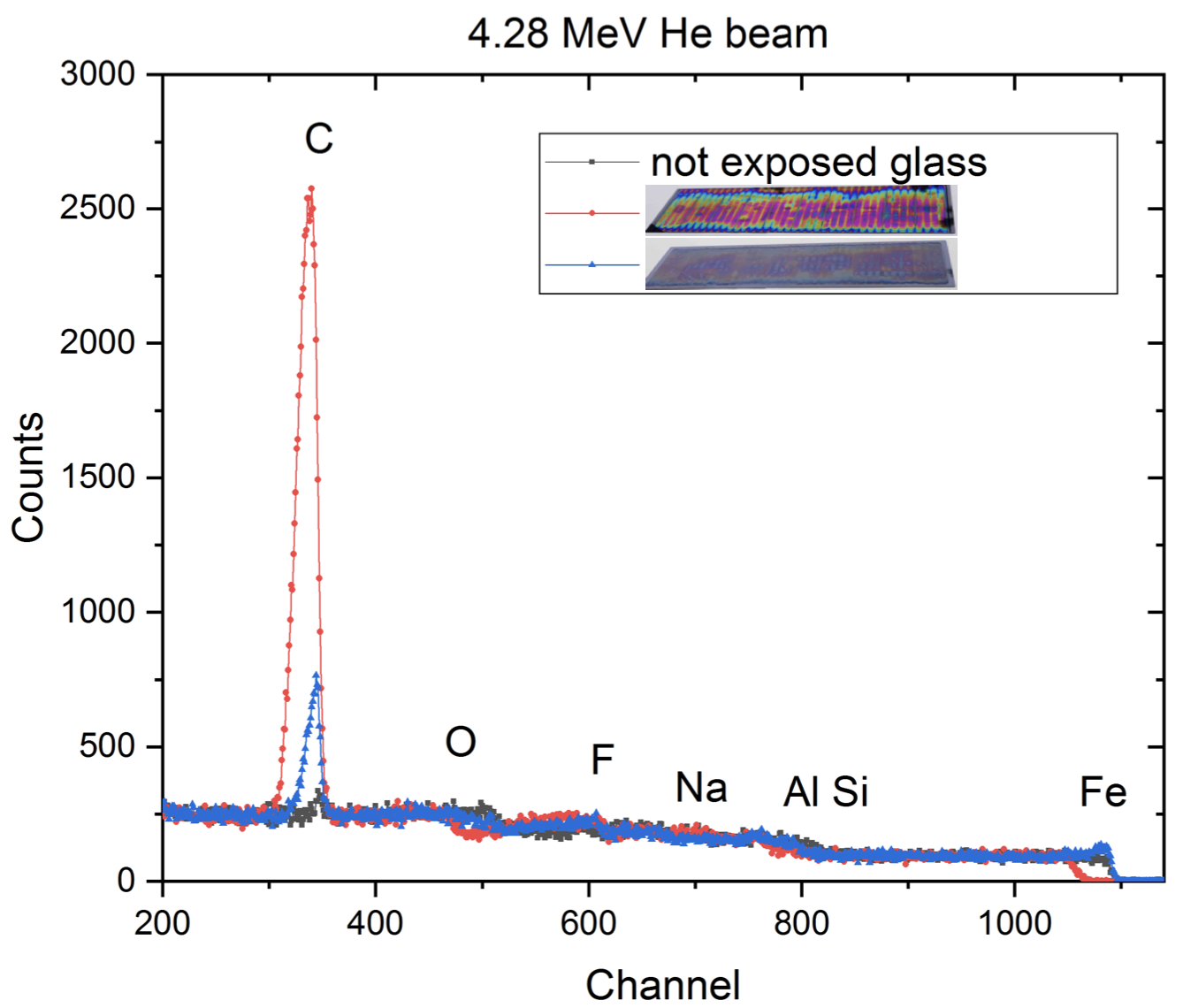}
	\caption{Overlapped $^{12}$C-$^4$He resonance spectra for: black colour - not 
	irradiated glass plate; blue - the cathode surface; red - the anode surface, respectively.}
	\label{FIG:23}
\end{figure}

 These results confirm the higher content in oxygen of the deposition on the cathode surface 
 and larger fluorine content on the anode surface. The missing
 Fe component on the cathode surface, which enters in the composition of the glass electrode 
 could be explained by the larger thickness of the deposited material.
 A precise carbon composition of the non-irradiated glass plate and the two surfaces of the irradiated one is obtained. The results are presented in 
 Fig.\ref{FIG:23} where the corresponding three spectra are overlapped. These results confirm that the largest carbon content is in the deposited layer on the anode surface, negligible amount being observed on the surface of non-irradiated glass. The uncertainty in RBS and foil-ERDA data analysis is about 5\% and 8\%, respectively.
  
\section{AFM analysis}

 Using atomic force microscopy (AFM), one could obtained information on the surface 
 roughness in three dimensions, giving information on the surface morphology
 \cite{bin, mol, burd}. In order to access a better resolution for the present analysis we
 used tapping mode. A MultiMode Nanoscope IIIA Controller was used for the measurements
 \cite{burd}. The measurements were done at the room temperature and a RTESP (Phosphorus (n) 
 doped Si) cantilever with an elasticity constant of 20-80 N/m. For data acquisition and off-line
 data processing AFM NanoScope v531r1 software environment was used.

\begin{figure}[h]
	\centering
		\includegraphics[scale=0.25]{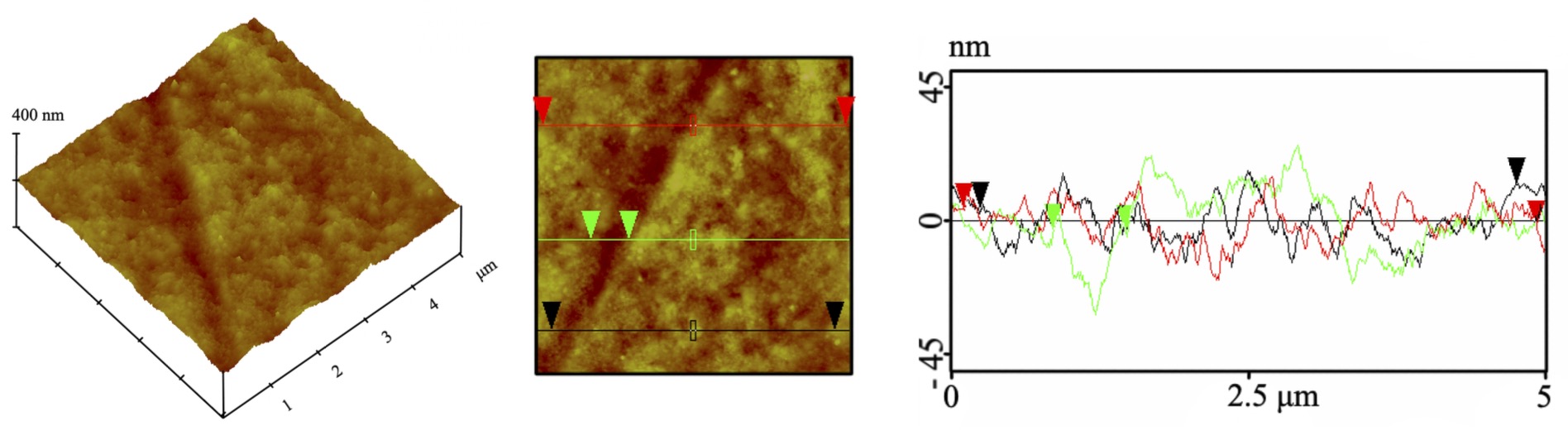}
	\caption{Left-3D representation of a 5x5$\mu$m$^2$ scanned surface of a non-irradiated
	 glass plate; Middle-2D projection of the 3D plot; Right-projections on one axis following
	  the three coloured lines shown in the middle plot.}
	\label{FIG:24}
\end{figure} 

\begin{figure}[h]
	\centering
		\includegraphics[scale=.5]{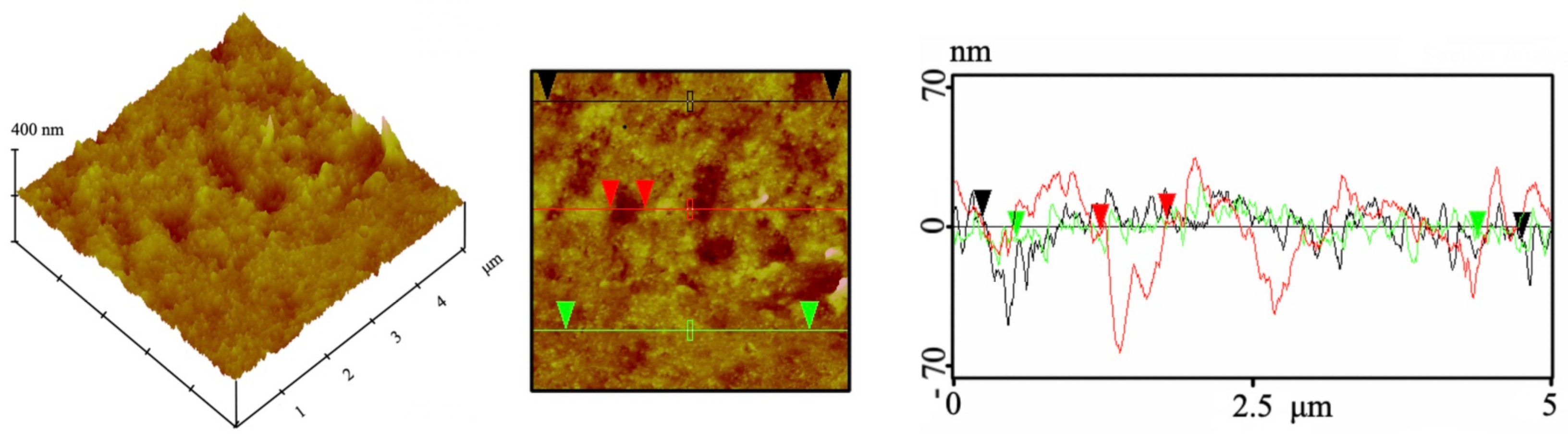}
	\caption{Left-3D representation of a 5x5$\mu$m$^2$ scanned surface of an irradiated
	 glass plate in the red-brick region of the anode surface; Middle-2D projection of the 3D 
	 plot; Right-projections on one axis following
	  the three coloured lines shown in the middle plot.}
	\label{FIG:25}
\end{figure} 

A 3D representation of a 5x5$\mu$m$^2$ scanned surface of a non-irradiated glass plate
is presented in Fig.\ref{FIG:24} left. Projections on one axis following the three 
coloured lines shown in middle of Fig.\ref{FIG:24} can be followed on the right plot. The observed structures with a peak to valley difference of 10-40 nm
are more or less at the same level as those reported in \cite{yw}.

\begin{figure}[h]
	\centering
		\includegraphics[scale=.48]{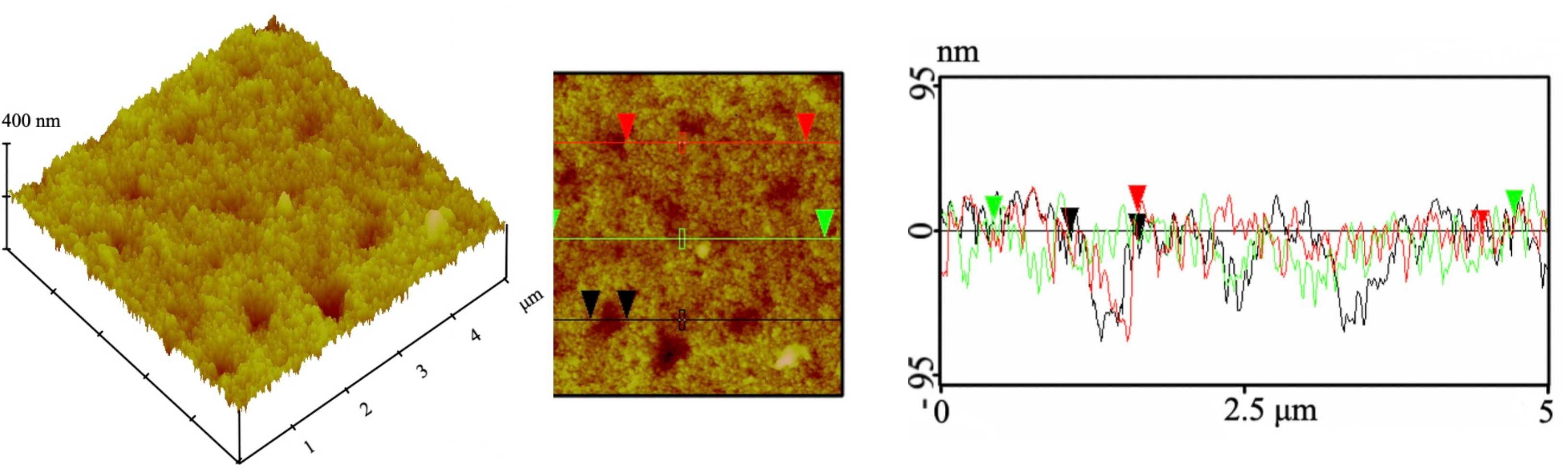}
	\caption{Left-3D representation of a 5x5$\mu$m$^2$ scanned surface of an
	irradiated
	 glass plate in the greenish region of the anode surface; Middle-2D projection of the 3D 
	 plot; Right-projections on one axis following
	  the three coloured lines shown in the middle plot.}
	\label{FIG:26}
\end{figure} 

\begin{figure}[h]
	\centering
		\includegraphics[scale=.48]{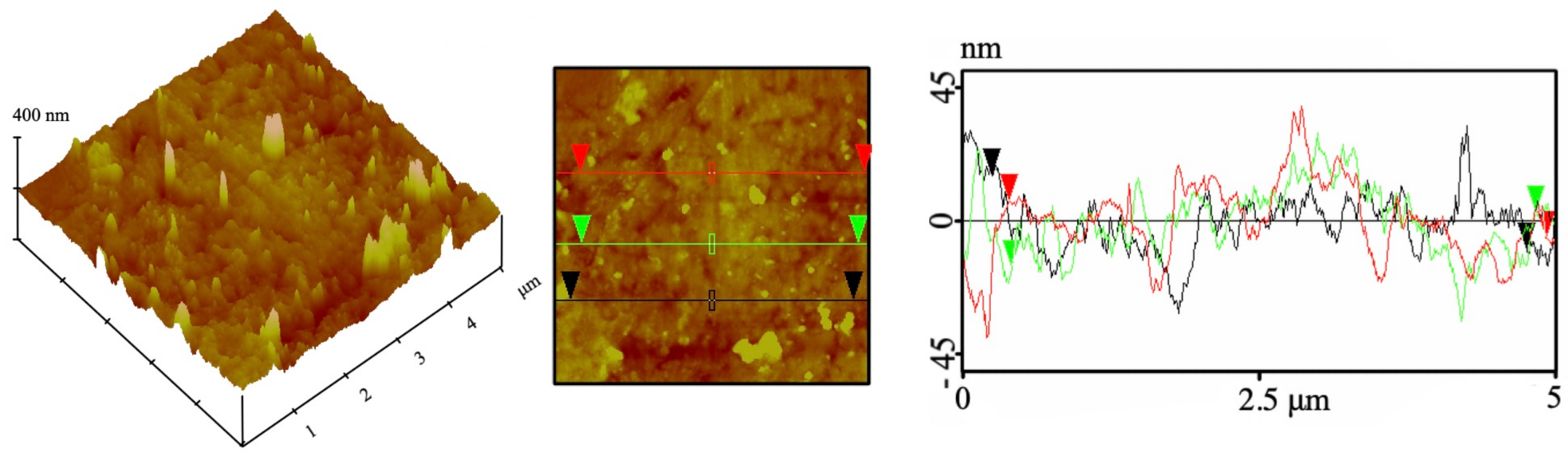}
	\caption{Left-3D representation of a 5x5$\mu$m$^2$ scanned cathode surface of an irradiated
	 glass plate; Middle-2D projection of the 3D 
	 plot; Right-projections on one axis following
	  the three coloured lines shown in the middle plot.}
	\label{FIG:27}
\end{figure}

 The results for a red-brick region on the anode surface of the exposed glass plate  are presented
 as 3D, 2D and uni-dimensional plots in Fig.\ref{FIG:25}.
 More pronounced structures are evidenced, the peak to valley differences being up to the 
 level of 100 nm.

%\begin{figure}[h]
%	\centering
%		\includegraphics[scale=.22]{}
%	\caption{3D representation of a 5x5$\mu$m$^2$ scanned surface of an irradiated glass plate
%	in greenish region of the surface facing the cathode.}
%	\label{FIG:28}
% \end{figure} 

 The results for a greenish region on the same side of the exposed glass plate are presented
 in Fig.\ref{FIG:26}.
 The peak to valley differences are similar with 
 those observed for the red-brick region.

%\begin{figure}[h]
%	\centering
%		\includegraphics[scale=.32]{Fig29.png}
%	\caption{Projections on one axis - top plot, following the three 
%coloured lines shown on bottom left side for an irradiated glass plate in the greenish region of the surface facing the cathode.}
%	\label{FIG:29}
%\end{figure}

 In Fig.\ref{FIG:27} are presented the results for the cathode surface of the irradiated glass 
 electrode. Sharp peaks, randomly distributed are evidenced,  
 the peak to valley differences being up to the level of 70-80 nm in the scanned region.

%\begin{figure}[h]
%	\centering
%		\includegraphics[scale=.22]{Fig30.png}
%	\caption{3D representation of a 5$\mu$m$^2$ scanned surface of an irradiated glass plate
%	facing the anode.}
%	\label{FIG:30}
%\end{figure} 

%\begin{figure}[h]
%	\centering
%		\includegraphics[scale=.32]{Fig31.png}
%	\caption{Projections on one axis - top plot, following the three 
%coloured lines shown on bottom left side for an irradiated glass plate - surface facing the anode.}
%	\label{FIG:31}
%\end{figure}

\section{THz-TDS analysis}

 As it is known, the terahertz time-domain spectroscopy (THz-TDS) is used to determine the
 optical and dielectric constants of a given material. Therefore, we used this method to 
 study the changes in the refractive index and real and imaginary dielectric constant of the 
 floating glass electrodes used in the MSMGRPC operated in high irradiation dose. 
\begin{figure}[h]
	\centering
		\includegraphics[scale=.4]{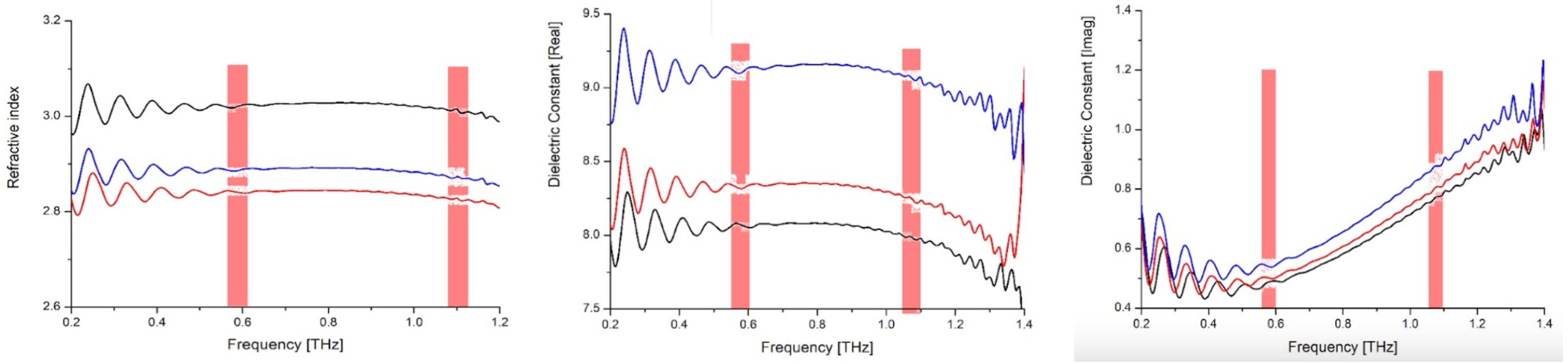}
	\caption{From left to right: refractive index, real part of dielectric constant and 
	imaginary part of the dielectric constant; black curves - not irradiated glass plate,
	red curves - the cathode surface and blue curves - anode surface. The measurement with the 
	highest accuracy is between the orange
	vertical stripes.}
	\label{FIG:28}
\end{figure}

 The results are presented in Fig.\ref{FIG:28}. A decrease of the refractive index 
 of $\approx$5.8\% is observed for the irradiated glass plate relative to the one not exposed
 to the radiation. The refractive index of the cathode layer
 is a bit lower relative to the anode layer. The dielectric constant, real and
 imaginary components are by $\approx$ 10\% lower for the surfaces of the exposed electrodes 
 relative to the values corresponding the 
 non-irradiated glass plate.

\begin{figure}[]
	\centering
		\includegraphics[scale=.45]{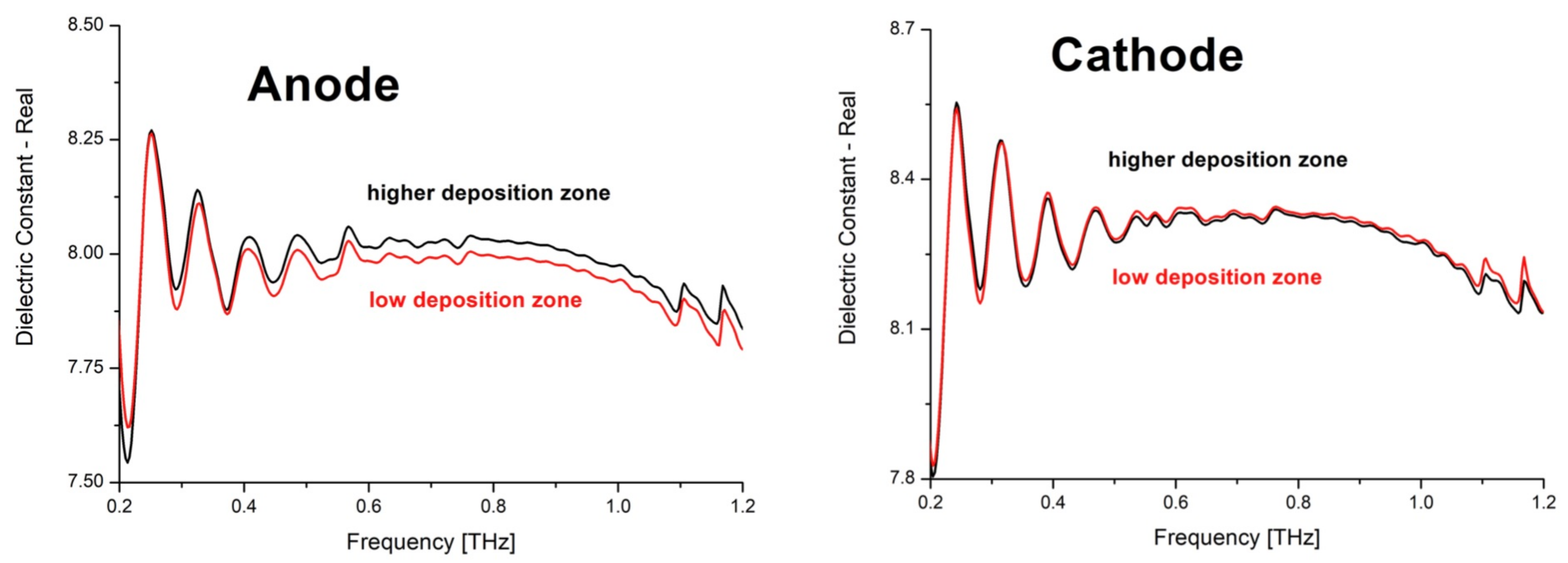}
	\caption{Left: real part of the dielectric constant for the anode surface of the glass electrode (Anode) in two regions: higher deposition and low deposition 
	zones. 
	    Right: real part of the dielectric constant for the cathode surface of the 
	    glass electrode (Cathode) in two regions: high deposition and low 
	    deposition. }
	\label{FIG:29}
\end{figure}

In Fig.\ref{FIG:29} are presented the variation of the real and imaginary parts of the dielectric constant with frequency corresponding to zones with different thickness of the deposited material on the two surfaces of a floating glass electrode. 
Based on the statistics of the thickness variation between low and high deposition areas, both for anode and cathode glass surfaces and correlating them with the dielectric constant variation presented in Fig.\ref{FIG:29}, one could have a 
qualitative assessment about the influence of the deposited layers on the electrode capacity variation. A capacity variation of about 0.05\% per deposited micrometer material is expected for the cathode surface and 0.7\% for the anode surface.

 \begin{figure}[b]
	\centering
		\includegraphics[scale=.4]{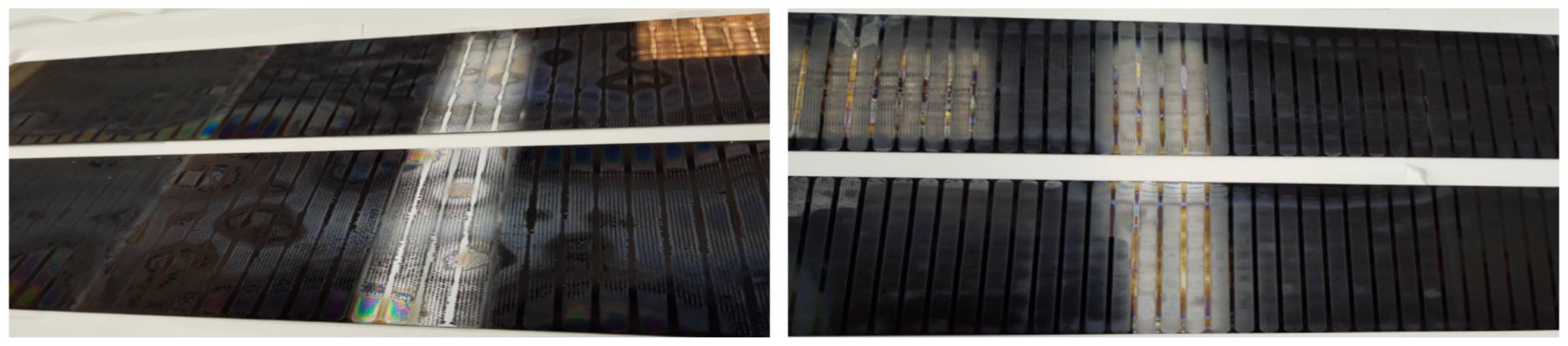}
	\caption{Photos of the two glass electrodes in touch with the strip structured high
voltage electrodes: Left: the surface in a direct contact with the cathode electrode; 
Right: the surface in direct contact with the anode electrode.}
	\label{FIG:30}
\end{figure}
 
 The observed
 changes in the dielectric constants of irradiated floating electrodes on the transmission
 line impedance is of the order of a few percent, therefore no major impact on the impedance
 matching to the one of the frontend electronics is expected.
      
\section{Resistivity measurements}

 An other important diagnose of the impact of high irradiation dose environment on the counter
 performance concerns the changes of the surface and volume resistivity of the exposed
 glass electrodes. These measurements were done for the floating glass electrodes and also 
 for the glass plates in direct contact with the strip structured high voltage electrodes.
 Photos of the surfaces of glass plates in a direct contact with the high voltage electrodes 
 are presented in Fig.\ref{FIG:30}.
 In general, the deposition seem to be less than on the floating glass electrodes, showing
 that most of the deposited material is due to ionisation and recombination of the basic
 components of the working gas in the gaps between consecutive glass electrodes. The layer on
 the surface in contact with the cathode seems to be thicker, regular patterns are visible and
 the edges of the gaps between strips are rather irregular relative to the ones produced on
 the surface in contact with the anode which are very sharp.

\begin{table}
\begin{tabular}{|c|c|c|}
\hline
Probe & $R_V$ (G$\Omega\cdot$cm) & $R_S$ ($G\Omega/\square$) \\
\hline
irradiated  &  &  \\
cathode surface & 67.4 & 20.0\\
\hline
irradiated anode &  &  \\
surface & 61.5 & 21.1 \\
\hline
non-irradiated glass & 65.2 & 20.2 \\
\hline           

\end{tabular}
\caption{Volume ($R_V$) and surface ($R_S$) resistivity for the two surfaces of
irradiate glass and for non-irradiated glass}
\end{table}

\begin{table}
\begin{tabular}{|c|c|c|c|c|}
\hline
Probe & $R_V$ (G$\Omega\cdot$cm) & $R_{SA}$ ($G\Omega/\square$) & $R_V$ (G$\Omega\cdot$cm) & $R_{SK}$ ($G\Omega/\square$)\\
\hline
1 & 64.6 & 19.1 & 62.6 & 20.8 \\
2 & 64.3 & 17.6 & 60.6 & 19.4 \\
3 & 64.0 & 16.8 & 58.5 & 17.5\\
4 & 62.1 & 17.9 & 61.5 & 17.3 \\
\hline           

\end{tabular}
\caption{Volume ($R_V$) and surface ($R_S$) resistivity for the two surfaces of
irradiate glass in contact with anode (SA) and cathode (SK) high voltage electrodes for 
four different regions on the surface.}
\end{table}

 The surface and volume resistivity for a floating glass electrode 
 and for the glass plates in contact with 
 the high voltage electrodes of irradiated MSMGRPC as well as for the non-irradiated glass 
 were measured 
 using a Keithley 6517B Electrometer with 8009 Test Fixture.  
 The results corresponding to floating glass electrodes used during irradiation and to a
 non-irradiated glass are presented in Table 4. If one considers the
 position dependence on surface, no major change in the surface and volume 
 resistivity between the irradiated and non-irradiated glass is observed.
 Similar measurements were performed for four samples taken from different regions of the
 glass plates in contact with the anode (A) and cathode (K) electrodes. The results
 presented in Table 5, show that, as in the case of floating electrodes, 
 the surface and volume resistivity are similar with the values corresponding to the 
 non-irradiated glass plates.
 
 \section{Conclusions}
 
  Detailed studies of the effects of high density avalanches induced by high
  irradiation dose in a short time on two-dimensional position sensitive timing
  resistive plate counters are reported. They confirm earlier studies on ablation and polymerisation taking place in low pressure plasma \cite{HY} or long term ageing studies in rather moderate irradiation dose \cite{SG}.
  Deposition of different radicals produced by polymerisation on the anode surface of the floating glass electrodes, with relative weights of different elements as a function of deepness of the deposited layer is evidenced. These layers could be easily cleaned using ethyl alcohol. Fluorine radicals which produce ablation/etching of the cathode surface of glass electrodes are evidenced. Besides a continuous layer, regular patterns are evidenced which could be explained as the results of already existing tiny patterns on the non-exposed glass which are enhanced by ablation/etching process. The surfaces roughness increases relative to the non-irradiated glass. Surface and volume resistivity, dark current, dark counting rate, efficiency and cluster size measurements after irradiation show that there is a recovery process which brings the counter to the initial performance.
  The present studies were performed using a housing box flashed with the working gas mixture. Therefore, the gas exchange in the 140 $\mu$m 10 gas gaps of the counter is via diffusion
  process. It is well known \cite{HY} that the polymerisation phenomena is inverse proportional with the gas flow. A directed gas flow through the gas gaps is expected to decrease the observed ageing phenomena. Such a new architecture is in progress to be realised. Based on the produced charge in the 
  counter during the operation in the high irradiation dose and the charge produced by a minimum ionising particle we estimated more than 3$\cdot$10$^5$ avalanches/$cm^2$$\cdot$sec, 10 times higher than the counting rate expected at CBM for the most forward region, i.e. low
  polar angles. Besides an increase of the gas flow through the gas gaps, an exposure to a 
  lower dose and longer time, conclusive for the operating conditions in CBM experiment, will 
  be performed.         
  
 \section{Acknowledgments}

This work was carried out under the contracts sponsored by the Romanian Ministry of Research, Innovation and Digitalization: 
CBM FAIR-RO-03 (via IFA Coordinating Agency), PN-19-06 01 03, PN-19-06 03 03, 
PN-19-06 02 01,
UEFISCU - CDI - HIGHkDEVICE, PN-III-P4-ID-PCCF-2016-0175.
The ion beam experiments were performed at the 3 MV Tandetron$^{TM}$ of IFIN-HH
being supported by the "National Interest Infrastructures" Program.

%\bibliographystyle{cas-model2-names}

%\bibliography{cas-refs}
%\bibliographystyle{cas-model2-names}
%\bibliography{cas-refs}
%\bibliography{test-refs}
%\bibliographystyle{}

\bibliographystyle{model6-num-names}
%\bibliographystyle{cas-model2-names}
%\bibliographystyle{elsarticle-num-names}

%\bibliography{refs.bib}
\bibliography{refs.blb}


\end{document}